\documentclass[aps,prl,preprint,superscriptaddress,longbibliography,nobibnotes]{revtex4-2}%\documentclass[aps,prl,twocolumn,superscriptaddress,longbibliography,nobibnotes]{revtex4-2}
\usepackage{diagbox}
\usepackage{amsmath}
\usepackage{amsfonts}
\usepackage{physics}
\usepackage[space]{grffile}
\usepackage{graphicx} % Include figure files
\usepackage{amssymb}
\usepackage{upgreek}
\usepackage{lineno}
\usepackage{natbib}
\usepackage{braket}
\usepackage{float}
\usepackage{color}
\usepackage{blindtext}
\usepackage{comment}
\usepackage{setspace}

\setcitestyle{super} %Make the citation as superscript

\usepackage{xr-hyper}
\usepackage{hyperref}
\usepackage{xcolor}
\hypersetup{colorlinks,breaklinks,
    urlcolor=[rgb]{0,0,0.64},
    linkcolor=[rgb]{0,0,0.64},
    citecolor=[rgb]{0,0,0.64},
    filecolor=[rgb]{0,0,0.64}}
\colorlet{BLACK}{black}

\usepackage[T1]{fontenc}

\newcommand{\MIT}{Massachusetts Institute of Technology, Department of Physics, Cambridge, Massachusetts 02139, USA}

\newcommand{\POSTECH}{Department of Physics, Pohang University of Science and Technology, Pohang 37673, Republic of Korea}
\newcommand{\SIMES}{SIMES, SLAC National Accelerator Laboratory, Menlo Park, California 94025, USA}
\newcommand{\SLAC}{SLAC National Accelerator Laboratory, Menlo Park, California, USA}
\newcommand{\BQI}{Beijing Academy of Quantum Information Sciences, Beijing 100913, China}
\newcommand{\CHESS}{CHESS, Cornell University, Ithaca, New York 14853, USA}
\newcommand{\MITEECS}{Massachusetts Institute of Technology, Department of Electrical Engineering and Computer Science, Cambridge, Massachusetts 02139, USA}
\newcommand{\PKU}{International Center for Quantum Materials, School of Physics, Peking University, Beijing 100871, China}
\newcommand{\UTokyo}{Department of Applied Physics, University of Tokyo, Bunkyo-ku, Tokyo 113-8656, Japan}
\newcommand{\TDL}{Tsung-Dao Lee Institute, School of Physics and Astronomy, and Zhangjiang Institute for Advanced Study, Shanghai Jiao Tong University, Shanghai 200240, China.}
\newcommand{\StanfordPAP}{Departments of Physics and of Applied Physics, Stanford University, Stanford, California 94305, USA}
\newcommand{\PAL}{Pohang Accelerator Laboratory, Pohang, Gyeongbuk 37673, Republic of Korea}

\begin{document}

\title{Dynamics of a jointly commensurate moiré charge density wave}
%TC:ignore

\author{Kyoung~Hun~Oh}
\thanks{These authors contributed equally to this work: K.H.O., Y.S., and H.N.}
\affiliation{\MIT}
\author{Yifan~Su}
\thanks{These authors contributed equally to this work: K.H.O., Y.S., and H.N.}
\email[Correspondence to: ]{ys3945@columbia.edu}
\affiliation{\MIT}
\author{Honglie~Ning}
\thanks{These authors contributed equally to this work: K.H.O., Y.S., and H.N.}
\affiliation{\MIT}
\author{B.~Q.~Lv}
\affiliation{\TDL}
\affiliation{\MIT}
\author{Alfred~Zong}
\affiliation{\StanfordPAP}
\affiliation{\SIMES}
\affiliation{\MIT}
\author{Dong~Wu}
\affiliation{\BQI}
\author{Qiaomei~Liu}
\affiliation{\PKU}
\author{Gyeongbo~Kang}
\affiliation{\PAL}
\author{Hyeongi~Choi}
\affiliation{\PAL}
\author{Hyun-Woo~J.~Kim}
\affiliation{\POSTECH}
\author{Seunghyeok~Ha}
\affiliation{\POSTECH}
\author{Jaehwon~Kim}
\affiliation{\POSTECH}
\author{Suchismita~Sarker}
\affiliation{\CHESS}
\author{Jacob~P.~C.~Ruff}
\affiliation{\CHESS}
\author{Xiaozhe~Shen}
\affiliation{\SLAC}
\author{Duan~Luo}
\affiliation{\SLAC}
\author{Stephen~Weathersby}
\affiliation{\SLAC}
\author{Patrick~Kramer}
\affiliation{\SLAC}
\author{Xinxin~Cheng}
\affiliation{\SLAC}
\author{Dongsung~Choi}
\affiliation{\MITEECS}
\author{Doron~Azoury}
\affiliation{\MIT}
\author{Masataka~Mogi}
\affiliation{\MIT}
\affiliation{\UTokyo}
\author{B.~J.~Kim}
\affiliation{\POSTECH}
\author{N.~L.~Wang}
\affiliation{\BQI}
\affiliation{\PKU}
\author{Hoyoung~Jang}
\affiliation{\PAL}
\author{Nuh~Gedik}
\email[Correspondence to: ]{gedik@mit.edu}
\affiliation{\MIT}

\date{\today}

\maketitle

\noindent\textbf{The advent of two-dimensional moiré systems has revolutionized the exploration of phenomena arising from strong correlations and nontrivial band topology. Recently, a moiré superstructure formed by two coexisting charge density waves (CDWs) with slightly mismatched wavevectors has been realized. These incommensurate CDWs can collectively exhibit commensurability, resulting in the jointly commensurate CDW (JC-CDW) and establishing a new paradigm for controlling moiré potential and periodicity. Achieving such functionality, however, hinges on a key open question: how do the amplitude, phase coherence, and periodicity of this order respond to external perturbations? Here, we address this question using a suite of time- and momentum-resolved diffraction and spectroscopic techniques to probe light-induced CDW dynamics in EuTe$_4$. Our time-resolved diffraction measurements distinguish the instantaneous quenching of the JC-CDW amplitude, as verified by time-resolved photoemission spectroscopy, from the much slower evolution of phase fluctuations. Furthermore, while the JC-CDW wavevector remains locked along the CDW direction upon photoexcitation, indicating a preserved moiré periodicity, the correlation length of JC-CDW shows an exclusive reduction perpendicular to its wavevector, unveiling the formation of previously unexplored shear-type defects. Together, this multimodal methodology reconstructs the spatiotemporal evolution of the JC-CDW upon excitation. These findings not only highlight the remarkable robustness of JC-CDWs out of equilibrium, but also provide insight into optical manipulation and engineering of moiré quantum materials through defect control.}\\

%TC:endignore
\section*{Introduction}

\begin{center}
\begin{figure}[H]
   \sbox0{\includegraphics[width=\textwidth]{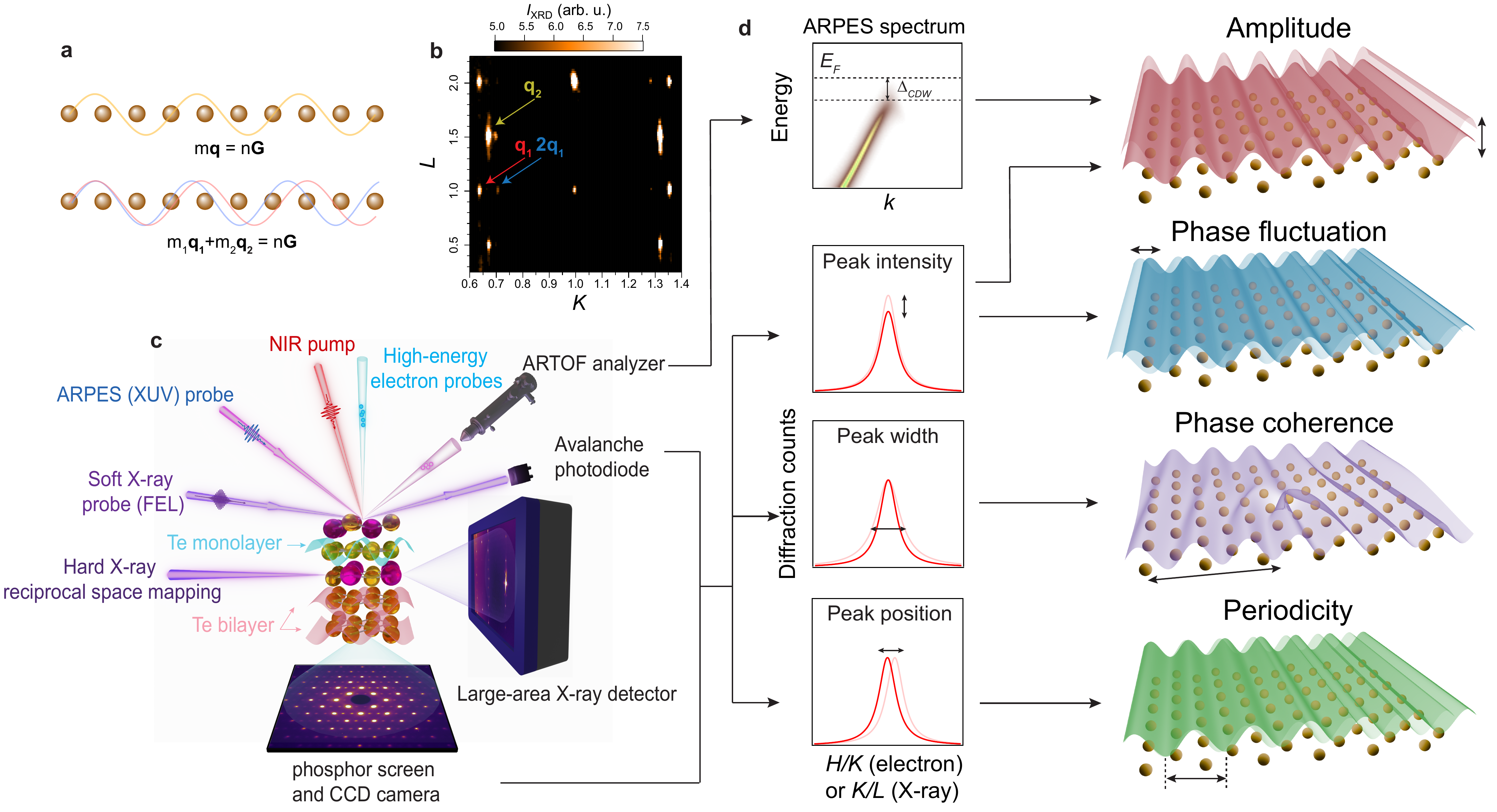}}% measure width
    \begin{minipage}{\wd0}
  \usebox0
  %\captionsetup{justification=raggedright,singlelinecheck=false}
  \caption{\textbf{\textcolor{black}{Workflow of time-and-momentum-resolved multimessenger probes for characterizing dynamics of joint commensurate charge density wave.}} \textbf{a} Schematics of a commensurate CDW (top panel) and jointly commensurate CDWs (bottom panel). Spheres represent the structural lattice and lines represent the charge density distribution. \textbf{b} Equilibrium XRD pattern in (2 $K$ $L$) plane. Red, blue, and gold arrows highlight the $\mathbf{\textcolor{black}{q_1}}$, $\mathbf{\textcolor{black}{2q_1}}$ and $\mathbf{\textcolor{black}{q_2}}$ CDW peaks, respectively. $I_{\text{XRD}}$ denotes the X-ray scattering intensity. \textbf{c} Schematic of the multi-messenger experimental setups, including static X-ray diffraction (XRD) in transmission geometry with a large-area detector, time-resolved soft X-ray diffraction (tr-XRD) in reflection geometry with an avalanche photodiode detector, time- and angle-resolved photoemission spectroscopy (tr-ARPES) with a time-of-flight detector, and ultrafast electron diffraction (UED) in transmission geometry with a phosphor screen and CCD detector. A near-infrared (NIR) pump is employed for all the time-resolved measurements. \textcolor{black}{\textbf{d} Physical quantities measured by each probe. The photoemission spectrum directly measures the CDW amplitude as reflected by the CDW band gap. Diffraction intensity is associated with both CDW amplitude and phase fluctuations. Diffraction peak width specifically reflects the correlation length of the CDW order. The peak position signifies the periodicity, and thus the wavevector of the CDW. Electron and X-ray diffraction measure the ($H$ $K$ 0) and (0 $K$ $L$) planes, respectively.}} 
  \label{fig:Fig1}
\end{minipage}
\end{figure}  
\end{center}

Two-dimensional van der Waals moiré systems have opened new avenues for exploring strongly correlated and topological physics, leading to exotic phases such as correlated insulators and unconventional superconductivity \cite{Cao2018UnconventionalSuperlattices, Cao2018CorrelatedSuperlattices, Mak2022SemiconductorMaterials}. 
Unlike conventional moiré systems that typically require artificial stacking, a natural, intrinsic moiré system was recently demonstrated in the newly synthesized layered material EuTe$_4$, where interference and reconstruction between two charge density waves (CDWs) with slightly mismatched wavevectors produce long-wavelength modulation \cite{Lv2024LargeWaves}.
Here, two incommensurate charge density waves (I-CDWs) jointly lock to the lattice, forming jointly commensurate CDWs (JC-CDWs) that bridge commensurate CDWs (C-CDWs) and I-CDWs by satisfying the relation $\sum_i m_{i} \mathbf{q_{i}} = n\mathbf{\hat{G}}$, where $\mathbf{q_{i}}$ are the CDW wavevectors of the constituent I-CDWs, $\mathbf{\hat{G}}$ is a unit reciprocal lattice vector, and $m_{i}$, $n$ are nonzero integers (Fig.~\ref{fig:Fig1}a) \cite{Emery1979Lockingsub3/sub,Lee1979ElectricWaves,Bruinsma1980Phase-locked/math, Ayari2004Sliding-inducedNbSe3}.
\textcolor{black}{Such a joint commensurability between coexisting CDWs has also been observed in NbSe$_3$, where the relationship between the wavevectors $\mathbf{q_{1}}$ and $\mathbf{q_{2}}$ remains robust upon application of an electric current \cite{Ayari2004Sliding-inducedNbSe3}.}

\paragraph{}
Notably, JC-CDWs exhibit emergent properties that define a new platform for exploring fundamental physics in a charge-ordered moiré superstructure.
The reconstructed long-period superlattices enable investigation of moiré phenomena in a bulk, scalable system with micron-scale coherence \cite{Lv2024LargeWaves}.
As the constituent I-CDWs can be responsive to perturbations, adjusting them allows engineering of the moiré potential and periodicity.
Additionally, since joint commensuration is typically satisfied along a specific crystallographic direction, JC-CDWs exhibit strong electronic directionality, offering opportunities to create new anisotropic spatial textures analogous to nematic or smectic electronic liquid crystal phases \cite{Kivelson1998ElectronicInsulator, Fradkin2010NematicPhysics}.
Lastly, the relative phase between individual I-CDWs introduces a novel degree of freedom, rendering an intrinsic energetic degeneracy among metastable states, which holds prospect for robust, non-volatile optical and electronic memory devices \cite{Lv2022UnconventionalWave,Wu2019LayeredSheets,Liu2024,Venturini2024}.

\paragraph{}
Understanding and controlling these properties of JC-CDWs requires a comprehensive characterization of their response against external stimuli.
Because charge orders are highly susceptible to optical excitation, their electronic states can be manipulated on ultrafast timescales, offering a pathway toward optical moiré engineering.
\textcolor{black}{In this context, recent studies have demonstrated photo-induced twist-angle engineering in moiré heterostructures \cite{Duncan2025PhotoinducedTwist}}
Yet, how the spatial configuration of JC-CDWs evolves under excitation remains largely unexplored.
In particular, it is unknown whether the \textcolor{black}{amplitude, phase fluctuations, spatial coherence, and periodicity} of each jointly locked I-CDWs exhibit enhanced robustness against perturbations compared to their uncoupled counterparts.
Since this stability is closely related to the nucleation and annihilation of topological defects --- features known to mediate ultrafast phase transitions in CDW systems \cite{Zong2018UltrafastWave, Geremew2019Bias-VoltageDevices, Zong2019EvidenceTransition,Kogar2020Light-inducedLaTe3,Zong2021RoleOrder,Cheng2024UltrafastWave,Ravnik2018,Vaskivskyi2015,Mihailovic2019TheApplications} --- it is plausible that JC-CDWs host unconventional defect dynamics, which further motivates comprehensive time-resolved characterization.

\paragraph{}
To address these challenges, we employ a suite of time- and momentum-resolved probes to monitor the dynamics of JC-CDWs following femtosecond laser excitation \cite{Zong2018UltrafastWave,Zong2021RoleOrder,Zong2019EvidenceTransition,Kogar2020Light-inducedLaTe3,Cheng2022Light-inducedCorrelations,Cheng2024UltrafastWave,Ravnik2018, Ning2024DynamicalSuperconductor,Su2025}.
Leveraging a combination of time-resolved X-ray diffraction (tr-XRD), \textcolor{black}{ultrafast electron diffraction (UED), and time- and angle-resolved photoemission spectroscopy (tr-ARPES), we separately track four key aspects of the CDW order parameter: amplitude, phase fluctuations, correlation length, and periodicity (Fig.~\ref{fig:Fig1}\textbf{c,d}). 
The CDW amplitude is directly reflected in the superlattice peak intensity in time-resolved diffraction and the gap opening near the Fermi surface in tr-ARPES. 
The CDW phase fluctuations can be retrieved from the CDW diffraction peak intensity, analogous to the Debye-Waller effect, in which incoherent phonons quench the Bragg peak intensity\cite{Overhauser1971ObservabilityDiffraction,Zong2019EvidenceTransition}. 
Furthermore, the spatial phase coherence is captured by the CDW diffraction peak width, which is inversely related to the correlation length\cite{Zong2019EvidenceTransition}. Finally, the CDW periodicity is obtained from the diffraction peak position, which directly determines the CDW wavevector. 
Collectively, these techniques provide a disentangled view of how different components of CDWs evolve out of equilibrium}, enabling three-dimensional, real-time visualization of JC-CDW and shedding light on its unexplored moiré dynamics.

\section*{\textcolor{black}{Results}}
\subsection*{\textcolor{black}{Structural characterization of JC-CDW}}
\paragraph{}
\textcolor{black}{Before studying the non-equilibrium dynamics of EuTe$_4$, we first confirm the JC-CDW structure of EuTe$_4$ by performing static X-ray diffraction (XRD) momentum space mapping at room temperature} (See Supplementary Note 1 for full diffraction images), well below its CDW transition temperature that is above 650 K \cite{Rathore2023EvolutionEuTe4}. 
The equilibrium XRD pattern presents a set of satellite peaks at $\mathbf{G} \pm \mathbf{\textcolor{black}{q_1}}$, $\mathbf{G} \pm \mathbf{\textcolor{black}{2q_1}}$, and $\mathbf{G} \pm \mathbf{\textcolor{black}{q_2}}$, where $\mathbf{G}$ is the reciprocal lattice vector, \textcolor{black}{$\mathbf{q_1}$=(0 0.644 0), and $\mathbf{\textcolor{black}{q_2}}$=(0 0.678 0.5)}, consistent with previous results (Fig.~\ref{fig:Fig1}b) \cite{Lv2022UnconventionalWave}.
Previous studies suggest that the peaks at $\mathbf{\textcolor{black}{q_1}}$ and $\mathbf{\textcolor{black}{q_2}}$ originate from two competing CDWs located in the monolayer and bilayer Te sheets, respectively, and are thus referred to as the monolayer CDW and bilayer CDW \cite{Lv2024LargeWaves, Ning2025}.
The small in-plane mismatch between $\mathbf{\textcolor{black}{q_1}}$ and $\mathbf{\textcolor{black}{q_2}}$ results in a long-wavelength modulation, corresponding to a moiré superstructure with a periodicity of $1/(\mathbf{\textcolor{black}{q_1}}-\mathbf{\textcolor{black}{q_2}})_{\mathrm{in-plane}}\sim$13.6 nm.
Consistent with prior findings, the relation $\mathbf{\textcolor{black}{q_1}}$+2$\mathbf{\textcolor{black}{q_2}}$=(0 2 1) holds, demonstrating these two CDWs form a JC-CDW even at room temperature\cite{Lv2024LargeWaves}.
Thus, EuTe$_4$ serves as an ideal testbed for investigating the properties of JC-CDWs.

\subsection*{\textcolor{black}{Temporal evolution of CDW order amplitude and phase fluctuations}}

\paragraph{}   
Having established the existence of the JC-CDW in EuTe$_4$ with steady-state measurements, we move on to investigate \textcolor{black}{the dynamics of the CDW order parameter amplitude and phase using tr-XRD.}
\textcolor{black}{While the intensity of a diffraction peak is often assumed to reflect only the CDW amplitude, it is also influenced by CDW phase fluctuations\cite{Zong2019EvidenceTransition,Overhauser1971ObservabilityDiffraction}. }
\textcolor{black}{In analogy to the Debye-Waller effect from incoherent phonons that suppress Bragg peak intensity, incoherent CDW phase fluctuations reduce the elastic CDW scattering intensity.
Therefore, transient changes in diffraction intensity generally represent a convolution of amplitude suppression and enhanced phase fluctuations, making their quantitative separation challenging.}

\begin{center}
\begin{figure}[H]
   \sbox0{\includegraphics[width=\textwidth]{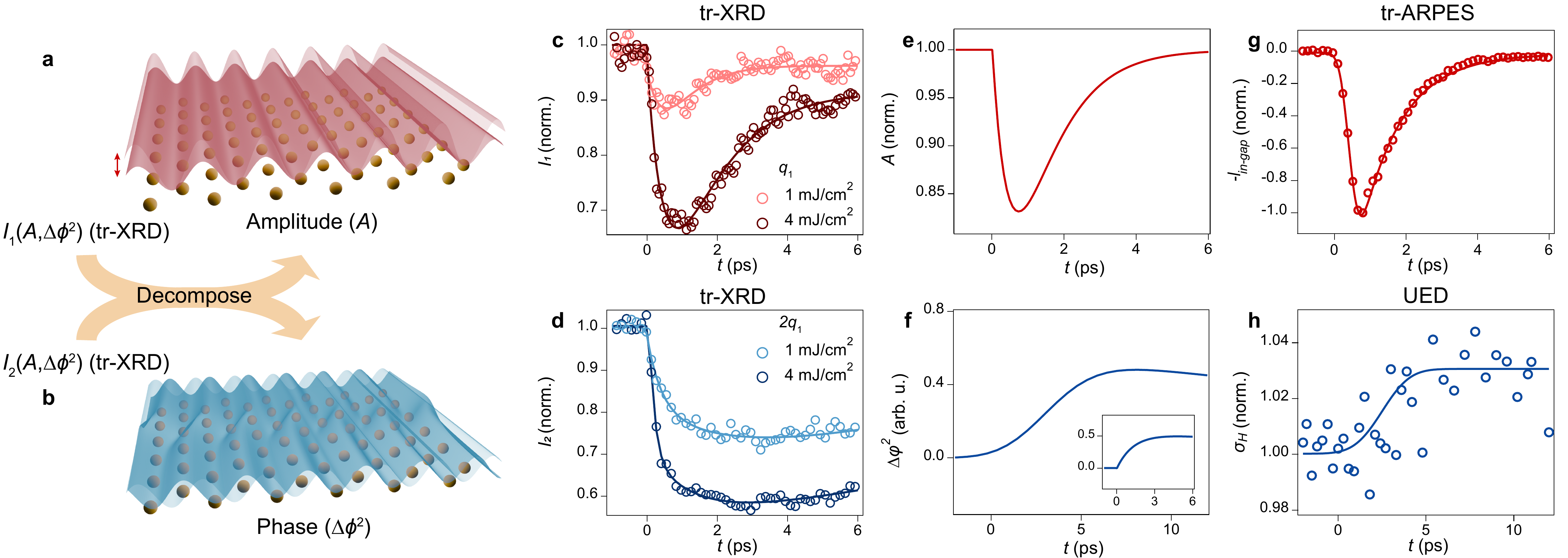}}% measure width
    \begin{minipage}{\wd0}
  \usebox0
  %\captionsetup{justification=raggedright,singlelinecheck=false}
  \caption{\textbf{\textcolor{black}{Decoding amplitude and phase fluctuations in the light-induced dynamics of the JC-CDW.}} \textbf{a-b} Schematics of \textbf{a} amplitude and \textbf{b} phase fluctuations. \textcolor{black}{The arrow denotes the workflow to extract amplitude and phase information from first and second order diffraction peaks in trXRD.}. \textbf{c} Temporal evolution of the intensity $I_1$ with pump fluences $F$ = 1 and 4 mJ/cm$^2$ of the CDW peak at $\mathbf{G}\pm\mathbf{\textcolor{black}{q_1}}$=(0 -0.644 1) normalized by its equilibrium value. Lines are fits to an exponential decay. \textbf{d} Temporal evolution of the intensity $I_2$ with pump fluences $F$ = 1 and 4 mJ/cm$^2$ of the CDW peak at $\mathbf{G}\pm\mathbf{\textcolor{black}{2q_1}}$=(0 0.712 1) normalized by its equilibrium value. Lines are fits to a bi-exponential decay. \textbf{e} \textcolor{black}{Extracted} temporal evolution of the monolayer CDW amplitude $A$, normalized to its equilibrium value, at a pump fluence of $F$ = 4 mJ/cm$^2$ \textcolor{black}{from tr-XRD (Supplementary Note 2).} \textbf{f} \textcolor{black}{Extracted} temporal evolution of the monolayer CDW phase fluctuations $\Delta \phi^2$ induced by light at a pump fluence of $F$ = 4 mJ/cm$^2$ \textcolor{black}{from tr-XRD (Supplementary Note 2). The result is further} convolved with the instrumental temporal resolution of UED ($\sim$2 ps) for direct comparison with \textbf{h}. Inset shows the original temporal evolution of $\Delta \phi^2$ without accounting for the cross-correlation function. \textbf{g} Temporal evolution of the negative in-gap photoelectron spectral intensity ($-I_{in-gap}$) extracted from tr-ARPES spectra taken with a pump fluence of $F$ = 0.8 mJ/cm$^2$, normalized to its maximal value around $\sim$0.8 ps. Solid line is a fit to a single exponential decay. \textbf{h} Temporal evolution of the diffraction peak width along the $H$ direction $\sigma_H$ with a pump fluences $F$ = 2.5 mJ/cm$^2$. Solid line is a fit to a single exponential decay.}
  \label{fig:Fig2}
\end{minipage}
\end{figure}  
\end{center}

\paragraph{}
\textcolor{black}{To disentangle these two effects,} we develop a previously overlooked analysis scheme by tracking the dynamics of the intensity of multiple CDW peaks corresponding to the same monolayer CDW order, namely the first-order ($\mathbf{\textcolor{black}{q_1}}$ peak) and second-order satellite peaks ($\mathbf{\textcolor{black}{2q_1}}$ peak).
This approach is justified as the amplitude (Fig.~\ref{fig:Fig2}a) and phase \textcolor{black}{fluctuation} (Fig.~\ref{fig:Fig2}b) modulations of the CDW contribute to changes in different satellite peak intensities to different extents\cite{Overhauser1971ObservabilityDiffraction}.
Quantitatively, the intensities of the first- and second-order peaks, $I_1$ and $I_2$, depend on the CDW amplitude $A$ and phase fluctuation $\Delta\phi^2$ as 
\begin{align}
\textcolor{black}{I_1} &\textcolor{black}{\sim A^2 \textup{e}^{-\frac{\Delta\phi^2}{2}}}\\
 \textcolor{black}{I_2} &\textcolor{black}{\sim (A^2 + \gamma A)^2 \textup{e}^{-2\Delta\phi^2},}
\end{align} 
where $\gamma$ is a constant determined by the CDW wavevector and shape (Supplementary Note 2).
Therefore, \textcolor{black}{this method allows us to extract} the dynamics of amplitude and phase fluctuations of the target CDW by simply measuring the intensity of multiple superlattice peaks without relying on \textcolor{black}{several dedicated experimental techniques}.

\paragraph{}
The intensity evolution of the $\mathbf{\textcolor{black}{q_1}}$ and $\mathbf{\textcolor{black}{2q_1}}$ peaks (Figs.~\ref{fig:Fig2}c,d) indeed exhibit stark differences: $I_1(t)$ rapidly decreases in $\sim1$ ps and shows nearly full recovery within $\sim6$ ps, while $I_2(t)$ slowly decreases until $\sim3$ ps and exhibits only slight recovery up to $\sim6$ ps.
This disparity remains consistent across the investigated pump fluences from 1 to 4 mJ/cm$^2$.
This observation strongly suggests that the dynamics of amplitude and phase fluctuations occur on different timescales following photoexcitation. 
We thus leverage the aforementioned method to unambiguously disentangle the temporal evolution of both components without any \textcolor{black}{\textit{a priori} assumptions} (Figs.~\ref{fig:Fig2}e,f, See Supplementary Note 2 for details).
The reproduced CDW amplitude decreases until $\sim0.8$ ps and nearly fully recovers in $\sim6$ ps (Fig.~\ref{fig:Fig2}e).
\textcolor{black}{Such rapid suppression and recovery timescales are consistent with the CDW dynamics previously reported in $R$Te$_3$} \cite{Zong2019EvidenceTransition}.
\textcolor{black}{By contrast, the extracted phase fluctuations} continue to increase until $\sim5$ ps and shows only partial recovery within the time window (Fig.~\ref{fig:Fig2}f).
This result aligns with the relatively slow time scales of phase dynamics previously reported in $R$Te$_3$ and nickelates \cite{Zong2019EvidenceTransition, Lee2012PhaseNickelate}.
Such observations also provide the first smoking-gun evidence of the significantly different formation and relaxation timescales of the amplitude and phase degrees of freedom of JC-CDWs.
Compared to the rapid nonthermal dynamics of the CDW satellite peaks, the structural Bragg peaks at (0 0 1) and (0 1 2) exhibit dynamics that are slower by an order of magnitude, which can be attributed to the thermal Debye-Waller effect (Supplementary Note 3)\cite{Chase2016UltrafastFilms,Lin2017UltrafastMoSe2,Tinnemann2019UltrafastTransfer}.

\paragraph{}
To verify the validity of the decomposition \textcolor{black}{result}, we employ tr-ARPES to directly track the CDW amplitude dynamics (Supplementary Note 4).
As previous static ARPES experiments have demonstrated CDW gap opening in EuTe$_4$ \cite{Lv2022UnconventionalWave}, we can use the in-gap spectral weight as a measure of the CDW gap size, which is proportional to the CDW amplitude \cite{Gruner2018DensitySolids,Sobota2021Angle-resolvedMaterials,Lv2022UnconventionalWave,Lv2024CoexistenceSemiconductor,Su2025,Zong2019EvidenceTransition}.
As demonstrated in Fig.~\ref{fig:Fig2}g, the negative in-gap intensity exhibits a rapid decrease up to $\sim$ 0.8 ps and a nearly full recovery by $\sim$ 6 ps, qualitatively resembling the temporal evolution of the \textcolor{black}{monolayer} CDW amplitude extracted from tr-XRD (Fig.~\ref{fig:Fig2}e).
Consequently, the tr-ARPES results successfully corroborate the validity of the CDW amplitude dynamics obtained from tr-XRD.

\subsection*{\textcolor{black}{Temporal evolution of CDW periodicity and correlation length}}

\begin{figure}[H]
\centering
\includegraphics[width=\textwidth]{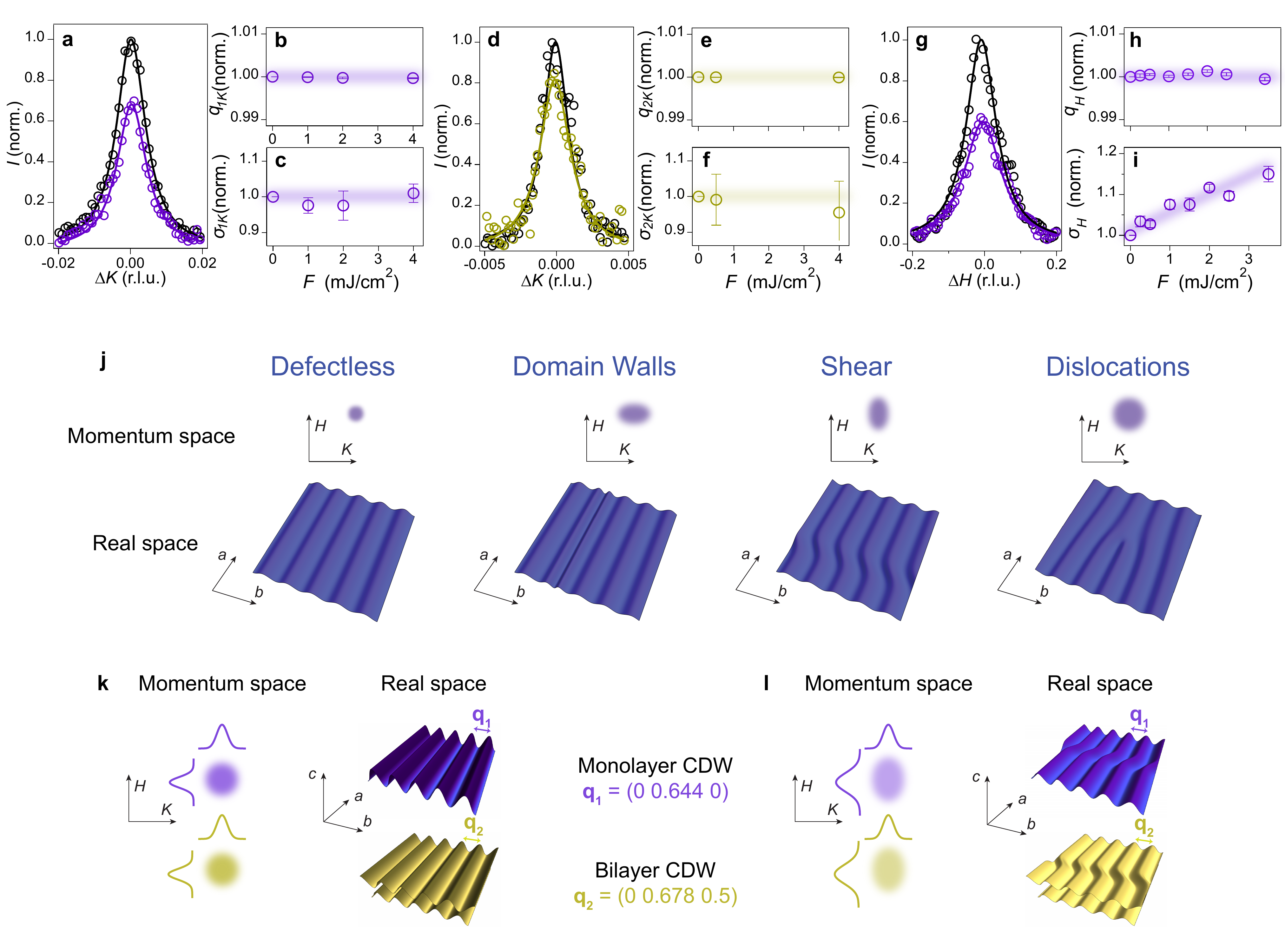}
\newpage
  \caption{\textbf{Light-induced changes in the JC-CDW periodicity and spatial coherence.} \textbf{a} $K$-cut of the $\mathbf{\textcolor{black}{q_1}}$ peak at (0 -0.644 1) taken at $t$= -2 ps (black) and 0.5 ps (purple) with $F$ = 4 mJ/cm$^2$ from tr-XRD. Lines are fits to Lorentzians. Error bars indicate the standard deviation of the fit. \textbf{b} Normalized position ($\textcolor{black}{q_{1K}}$) and \textbf{c} width ($\sigma_{K}$) defined as the full width at half maximum (FWHM) of the $\mathbf{\textcolor{black}{q_1}}$ peak, taken at 0.5 ps. Shaded lines are guides to the eyes. \textbf{d} $K$-cut of the $\mathbf{\textcolor{black}{q_2}}$ peak at (0 0.678 1.5) taken at $t$= -2 ps (black) and 1 ps (yellow) with $F$ = 4 mJ/cm$^2$ from tr-XRD. Lines are fits to Lorentzians. Error bars indicate the standard deviation of the fit. \textbf{e} Normalized peak position ($\textcolor{black}{q_{2K}}$) and \textbf{f} width ($\sigma'_{K}$) of the $\mathbf{\textcolor{black}{q_2}}$ peak, taken at 1 ps. Shaded lines are guides to the eyes. \textbf{g} $H$-cut of the $\mathbf{\textcolor{black}{q_1}}$ peak at (3 0.356 0) taken at $t$= -5 ps (black) and 4 ps (purple) with $F$ = 4 mJ/cm$^2$ from UED. Lines are fits to Lorentzians. Error bars indicate the standard deviation of the fit. \textbf{h} Normalized peak position ($q_{H}$) and \textbf{i} width ($\sigma_{H}$) of the $\mathbf{\textcolor{black}{q_1}}$ peak along $H$, taken at 4 ps. Shaded lines are guides to the eyes. \textcolor{black}{\textbf{j}. Real and momentum space schematics of three types of CDW defects. Compared to the defect-less ground state (left panel), the three types of defects --- domain walls, shear defects, and dislocations --- feature longitudinal, transverse, and isotropic broadening in CDW peak widths with respect to the CDW wavevector, respectively. } \textbf{k-l} Schematics of the \textbf{k} equilibrium and \textbf{l} excited diffraction peak shape for the monolayer CDW peak ($\mathbf{\textcolor{black}{q_1}}$ peak, purple) and the bilayer CDW peak ($\mathbf{\textcolor{black}{q_2}}$ peak, yellow). Purple (yellow) wavy patterns represent the corresponding real-space CDW configurations of the monolayer (bilayer) CDW. The intensity of the color indicates the amplitude of the order parameter. Purple (yellow) lines denote the line cuts of the CDW peaks along the $H$ and $K$ directions. }
\label{fig:Fig3}
\end{figure}

\paragraph{}
\textcolor{black}{As we have quantitatively identified the distinct timescales of the CDW amplitude and phase fluctuation dynamics, we now turn to understanding the real-space structural evolution of the JC-CDW by tracking the CDW superlattice peak profile along the $H$, $K$, and $L$ directions using time-resolved diffraction.
We first examine the peak profile along the CDW wavevector direction $K$ under strong photoexcitation that melts the CDW }\cite{Rischel1997FemtosecondFilms,Beaud2014ATransitions,Ning2024DynamicalSuperconductor}. 
\textcolor{black}{Notably, we observe} no discernible change in both the monolayer (Figs.~\ref{fig:Fig3}a,b) and bilayer (Figs.~\ref{fig:Fig3}d,e) CDW wavevectors at the time delay when the peak intensities reach their maximal suppression.
We note that this apparent invariance cannot be attributed to trivial experimental artifacts, such as pump–probe penetration-depth mismatch (Supplementary Note 5), nor to a delayed response that would emerge only at later time delays (Supplementary Note 6).
This absence of peak position change reflects the persistent wavevector-locking relation between the monolayer CDW and the bilayer CDW, $\textcolor{black}{q_{1K}}$+2$\textcolor{black}{q_{2K}}$=2 even out of equilibrium.
This behavior contrasts with typical I-CDWs, such as the dominant $c$-axis CDW in $R$Te$_3$ systems that share a structural motif similar to EuTe$_4$, whose wavevectors are susceptible to both temperature and light excitation \cite{Banerjee2013ChargeTellurides, Moore2016Ultrafast3, Blanco-Canosa2014ResonantX, Jang2016IdealYBCO, Jacques2016Laser-InducedChromium,Zhou2021NonequilibriumWave}.
Moreover, the peak widths along the wavevector direction of both CDWs do not broaden against light excitation either (Figs.~\ref{fig:Fig3}c,f).
\textcolor{black}{This observation seemingly contradicts the peak broadening frequently observed in various CDW materials, which is attributed to the optical generation of topological defects that reduce the correlation length}  \cite{Zong2019EvidenceTransition,Vogelgesang2018PhaseDiffraction}.
The invariant peak width upon photoexcitation indicates that the \textcolor{black}{CDW} defects that disrupt the \textcolor{black}{correlation length} along the CDW wavevector direction \textcolor{black}{are not generated}.
The robustness of peak position and width upon excitation jointly highlights the persistence of joint commensuration even out of equilibrium, reminiscent of the stability of C-CDWs.
\textcolor{black}{Analogously, we also observe invariance of peak position and width along $L$ (Supplementary Note 7), indicating the persistence of joint commensuration in the entire (0 $K$ $L$) plane where the CDW wavevectors reside.}

\paragraph{}
\textcolor{black}{We then} track the temporal evolution of the peak width along $H$, the other in-plane direction perpendicular to the CDW wavevectors.
Although tr-XRD can unambiguously probe CDW peaks with high momentum resolution, the typical tr-XRD experimental geometry employing a two-circle diffractometer can only access peaks in the (0 $K$ $L$) plane and prevent the detection of CDW shape changes along the $H$ direction.
To overcome these limitations, we utilize UED in transmission geometry, which enables time-resolved probing of peaks in the entire ($H$ $K$ 0) plane (Supplementary Note 8 for the full UED pattern).
\textcolor{black}{Since the bilayer CDW diffraction peaks are at half-integer $L$ values and are therefore undetectable by UED}, we henceforth focus on the monolayer CDW.
We present the linecut of the representative monolayer CDW peak along $H$ upon light excitation (Fig.~\ref{fig:Fig3}g).
Notably, the peak position exhibits no significant change upon photoexcitation up to 3.5 mJ/cm$^2$ (Figs.~\ref{fig:Fig3}h), reflecting the locking of CDW wavevector along $H$ akin to the $K$ and $L$ directions.
However, in contrast to them, the peak width along $H$ shows a clear increase, a hallmark of the formation of \textcolor{black}{light-induced defects that reduces the CDW correlation length.
We note that the transverse transient broadening of the CDW diffraction peak is unlikely to originate from convention structural disorder, in-plane hopping anisotropy of the Te $5p$ orbitals, or anisotropy in the equilibrium phase stiffness along the two in-plane directions (Supplementary Note 9).}
As we increase the pump fluence, the CDW peak width along $H$ increases accordingly (Fig.~\ref{fig:Fig3}i), suggesting the progressively increasing density of defects.

\paragraph{}
The exclusive peak width broadening of the monolayer CDW peak along the $H$ direction indicates that light-induced \textcolor{black}{CDW} defects disrupt the CDW coherence selectively along $H$ while preserving it along $K$ and $L$ (Figs.~\ref{fig:Fig3}j,k).
This behavior of JC-CDW contrasts strikingly with both previously studied photoexcited I-CDWs, such as $R$Te$_3$, and C-CDW systems, such as TiSe$_2$: 
in $R$Te$_3$, dislocation-type topological defects isotropically broaden the CDW diffraction peak along both in-plane directions \cite{Zong2019EvidenceTransition,Zhou2021NonequilibriumWave}, whereas in TiSe$_2$, 1D domain wall-like defects primarily broaden the CDW peak width along the wavevector direction (Fig.~\ref{fig:Fig3}j) \cite{Cheng2024UltrafastWave}.
Our observation echoes a hitherto rarely discussed shear-like defects, which form at the boundary between coherent CDW patches with a phase slip along the direction of the CDW wavevector (Fig.~\ref{fig:Fig3}j).
Therefore, these defects exclusively diminish the CDW \textcolor{black}{correlation length} along the in-plane direction perpendicular to the wavevector while preserving it along the wavevector (Figs.~\ref{fig:Fig3}k,l).
Consequently, our findings reveal the photo-induced formation of a new type of shear-like defects in EuTe$_4$ CDWs, underlining a unique inequilibrium property of JC-CDWs.
This behavior may stem from the exceptionally robust wavevector-locking of the two incommensurate CDWs forming the JC-CDW, which persists even out of equilibrium.
\textcolor{black}{We note that previous studies have reported a phenomenologically similar effect featuring CDW wavefront bending induced by a direct current, which also results in transverse broadening of the CDW diffraction peak. 
However, given the fundamental difference in generation mechanism, spatial scale, and structural requirement, such CDW bending is unlikely to be the primary origin of the anisotropic broadening observed in EuTe$_4$ (Supplementary Note 9).}

\paragraph{}
\textcolor{black}{Last, we track the temporal evolution of the correlation length by measuring the in-plane monolayer CDW peak width along $H$ with UED, the only direction that exhibits a light-induced change (Supplementary Note 10 for details).
We observe} a slow increase over $\sim$ 4 ps and a remarkably negligible recovery within the $\sim$ 12 ps time window (Fig.~\ref{fig:Fig2}h). 
Importantly, this behavior qualitatively resembles the temporal evolution of the CDW phase fluctuations extracted from tr-XRD (Fig.~\ref{fig:Fig2}f), after we take account of the order-of-magnitude lower temporal resolution of UED compared to tr-XRD.
We note that CDW phase fluctuations, which are Debye-Waller-like vibrations of the electronic structure that reduce satellite peak intensity, and phase decoherence, which is the loss of spatial phase correlation that broadens the satellite peak width, are not necessarily proportional quantities during an ultrafast photoexcitation process \cite{Lee2012PhaseNickelate}.
Nevertheless, their similar temporal evolution suggests a direct connection between these two quantities in the EuTe$_4$ CDW. 
This implies that the light-induced phase decoherence of CDW may directly result from the light-induced phase fluctuations, or vice versa.

\section*{\textcolor{black}{Discussion}}
\begin{center}
\begin{figure}[H]
   \sbox0{\includegraphics[width=\textwidth]{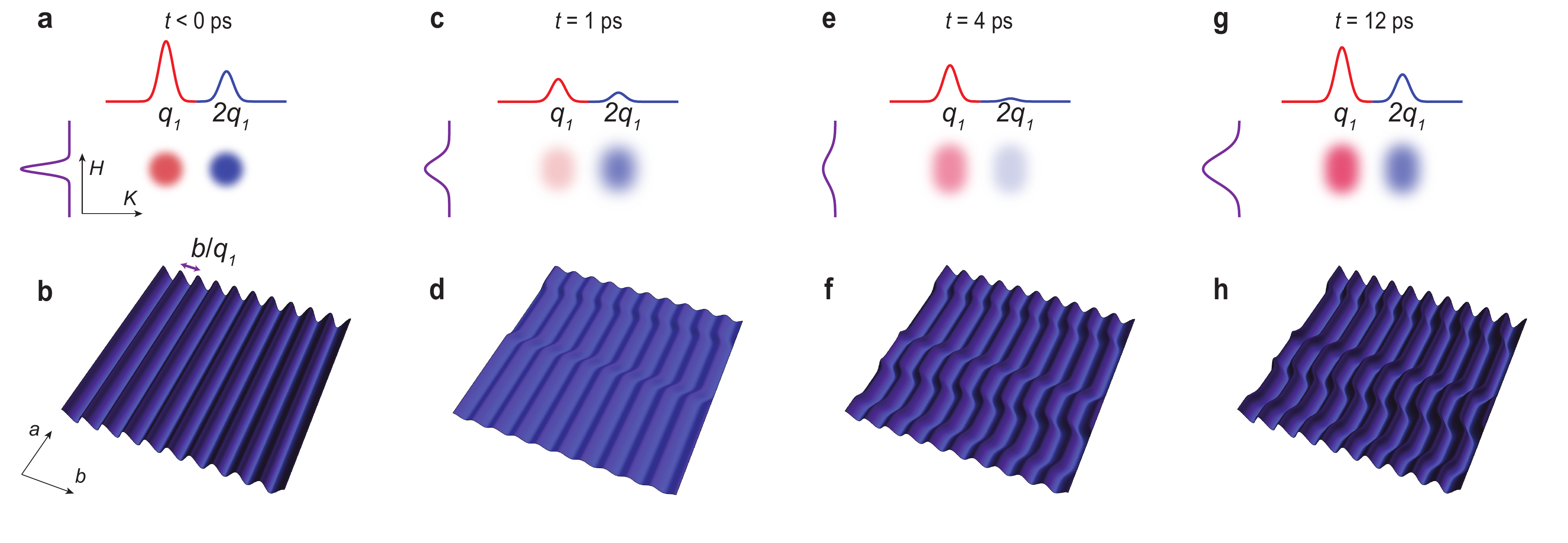}}% measure width
    \begin{minipage}{\wd0}
  \usebox0
  %\captionsetup{justification=raggedright,singlelinecheck=false}
  \caption{\textbf{Temporal evolution of the CDW spatial configuration and shear defects.} Schematics of the temporal evolution of the first- and second-order monolayer CDW peaks in momentum space and the corresponding real-space structure, at \textbf{a,b} $t$ < 0 ps, \textbf{c,d} $t$ = 1 ps, \textbf{e,f} $t$ = 4 ps, and \textbf{g,h} $t$ = 12 ps. Red (blue) lines represent the line cuts of the peak along the $K$ direction at $\mathbf{\textcolor{black}{q_1}}$ ($\mathbf{\textcolor{black}{2q_1}}$). Purple lines represent the line cut of the peak along the $H$ direction. Red (blue) ovals represent the shape of the $\mathbf{\textcolor{black}{q_1}}$ ($\mathbf{\textcolor{black}{2q_1}}$) satellite peaks. Purple wavy patterns represent the real-space CDW configurations of the monolayer CDW. The intensity of the color represent the amplitude of the order parameter. } 
  \label{fig:Fig4}
\end{minipage}
\end{figure}  
\end{center}

\paragraph{}
% Physical picture
\textcolor{black}{With the multimodal approach combining tr-XRD, UED, and tr-ARPES that offers outstanding momentum- and time-resolutions, we summarize the dynamics of the monolayer CDW configuration as follows} (Figs.~\ref{fig:Fig4}a,b).
Upon light irradiation, the amplitude of the monolayer CDW transiently decreases, accompanied by the formation of shear-type defects. 
At $\sim$1 ps, it is maximally quenched (Fig.~\ref{fig:Fig4}c,d), since both the the in-gap spectral weight and the CDW order amplitude extracted from tr-XRD intensity show a maximal decrease.
Simultaneously, the density of shear-type defects starts to grow, as indicated by the exclusive broadening of the CDW peak along the $H$ direction.
Around $\sim$4 ps, the amplitude of the monolayer CDW has mostly recovered (Fig.~\ref{fig:Fig4}e,f), while the degree of phase fluctuations is maximized.  
This is accompanied by a peak density of shear-type defects that decrease the \textcolor{black}{correlation length} of the monolayer CDW.
Finally, the total density of defects and CDW phase fluctuations remain nearly unchanged over $\sim$12 ps, while the amplitude of CDW has fully recovered to equilibrium (Fig.~\ref{fig:Fig4}g,h).

\paragraph{}
As the JC-CDW in EuTe$_4$ forms a long-period moiré superstructure \cite{Lv2024LargeWaves}, our observations of its light-induced dynamics show hallmarks of optical control of moiré systems. 
The transient suppression of the JC-CDW amplitude reflects a quenching of the moiré potential, as a reduced CDW amplitude in each layer weakens the total electrostatic potential.
By contrast, the JC-CDW periodicity remains unchanged, indicating that the moiré wavelength $\sim1/(\mathbf{\textcolor{black}{q_1}}-\mathbf{\textcolor{black}{q_2}})_{\mathrm{in-plane}}$ should be preserved.
This light-induced modification is distinct from conventional moiré engineering --- such as varying the twist angle in twisted multilayer systems or tuning the lattice constant in lattice-mismatched heterostructures --- which alter both the moiré periodicity and potential depth simultaneously.
Our results demonstrate that ultrafast optical excitation can potentially suppress the depth of the moiré potential selectively while maintaining its long-range periodicity, offering a new efficient route to tune moiré electronic landscapes.

\paragraph{}
% Summary and outlook. 
In summary, with a suite of time- and momentum-resolved techniques, we investigated the ultrafast light-induced dynamics of the JC-CDW phase in EuTe$_4$.
Light-induced changes in the diffraction peak position and width, investigated through a combination of tr-XRD and UED, demonstrate that the periodicity and \textcolor{black}{correlation length} of the JC-CDW along its wavevector direction remain invariant under strong light excitation, resembling the robustness of a C-CDW rather than an I-CDW.
By contrast, the \textcolor{black}{correlation length} of the monolayer CDW along the $H$ direction is largely disrupted, signifying the ultrafast formation of shear-type defects enabled by the exceptional robustness of the wavevector-locking of the JC-CDW. 
On the other hand, our tr-XRD measurements unambiguously reveal the disentangled temporal evolution of the amplitude and phase fluctuations of the monolayer CDW in EuTe$_4$ upon photoexcitation, achieved by simultaneously tracking the intensities of the first- and second-order diffraction peaks. 
Synergetic tr-ARPES and UED measurements validate this approach by revealing the distinct timescales of the melting and recovery dynamics for the CDW amplitude and phase fluctuations.
The invariance of the CDW periodicity, together with its amplitude suppression, indicates that the moiré periodicity is preserved while the moiré potential depth is reduced.
Our findings not only establish a novel approach for momentum- and time- resolved characterization of the CDW order parameter, but also constitute the first experimental observation of JC-CDW dynamics out of equilibrium.
Moreover, the novel dynamical methodology for distinguishing the amplitude and phase dynamics from higher-order diffraction peaks can be applied to other CDW systems where the higher-order peak intensity is strong enough.
This work opens new pathways not only for optical engineering of moiré systems but also for controlling \textcolor{black}{CDW} defects via ultrafast optical manipulation, offering prospects for uncovering emergent phases in quantum materials and enabling potential CDW-based nonvolatile memory applications \cite{Wang2019WritingPulses, Mihailovic2019TheApplications,Liu2024,Venturini2024}.

%TC:ignore
\section*{Methods}

\paragraph{Sample preparation\\} 
EuTe$_4$ single crystals were grown via a solid-phase reaction with Te as the flux.
A detailed description of the synthesis process can be found elsewhere \cite{Wu2019LayeredSheets, Lv2022UnconventionalWave}. 
For synchrotron-based XRD measurements, EuTe$_4$ flakes of approximately 100~nm thick were acquired by mechanical exfoliation and transferred to amorphous glass substrates. 
For tr-XRD measurements, EuTe$_4$ bulk crystals with a dimension of approximately 2$\times$1$\times$0.2 mm were used.
A fresh and flat surface was acquired by mechanical exfoliation before the measurement.
EuTe$_4$ thin flakes with a dimension of approximately 500$\upmu$m $\times$ 250$\upmu$m $\times$ 50 nm were mechanically exfoliated from bulk crystals and then transferred onto a copper TEM grid for the keV-UED experiments. 
In all measurements, the samples were mounted with their surface perpendicular to the crystalline $c$-axis.
No signs of degradation were observed throughout the sample preparation and measurements.
All experiments were conducted at room temperature.

\paragraph{Static XRD and ED experiment\\}
Static X-ray diffraction measurements were carried out at the Cornell High Energy Synchrotron Source (CHESS) Beamline 4B using a 29~keV beam.
High-throughput volumetric reciprocal space mapping was performed in transmission mode, where the sample was rotated around an in-plane axis. 
The diffracted signal was collected by a Dectris Pilatus 6M detector. 
The ED image shown in Fig.~\ref{fig:Fig1}b is taken at the MeV-UED beamline at SLAC National Accelerator Laboratory \cite{Weathersby2015Mega-electron-voltLaboratory}.
The probe pulse was frequency-tripled to generate 4.2~MeV electron bunches via photoemission from a copper target. 
The photoelectrons were then accelerated in a radio-frequency photo-injector at a repetition rate of 360~Hz. 
The diffraction pattern was imaged by a phosphor screen (P-43) and recorded by an electron-multiplying charge-coupled device (EMCCD, Andor iXon Ultra 888). 
A circular through-hole at the center of the phosphor screen allowed the undiffracted electron beam to pass through, preventing camera saturation.

\paragraph{Tr-XRD experiment\\}
All tr-XRD experiments were conducted at the SSS-RSXS endstation of PAL-XFEL \cite{Jang2020Time-resolvedLaser}. 
The probe was a time-delayed, linearly $\pi$-polarized soft X-ray pulse with an 80 fs duration and a 60 Hz repetition rate.
The pump was a s-polarized near-infrared (NIR) pulse, centered at 800 nm and generated by a Ti:Sapphire laser, with a 30 Hz repetition rate and pulse duration of 50 fs. 
The pump and probe beams were nearly collinear, with an angular separation of less than 1$^\circ$.
The probe energy, selected to avoid resonance with the absorption edges of Eu and Te, was set to 988 eV, 1226 eV, and 1050 eV for measuring peaks at (0 -0.644 1) , (0 0.678 1.5), and (0 0.712 1), respectively.
The spot size of the X-ray was approximately $100\times150$ $\upmu$m$^2$ (FWHM) at the sample position, while the spot size of the 800~nm pump laser was approximately $500\times500$ $\upmu$m (FWHM), assuming normal incidence. 
The X-ray scattering signal was detected using an avalanche photodiode, shielded with a 300-nm-thick aluminum filter to ensure light tightness. 
Data were recorded on a shot-to-shot basis using high-speed digitizers.

\paragraph{UED experiment\\}
keV-UED experiments were performed with the home-built UED system at MIT \cite{Freelon2023DesignInstrument}. 
The 1038~nm (1.19~eV) output of a commercial Yb:KGW regenerative amplifier laser system (PHAROS SP-10-600-PP, Light Conversion),  operating at a repetition rate of 5~kHz, was split into pump and probe branches. 
The pump branch was focused onto the sample for photoexcitation \textcolor{black}{(s-polarization)}, while the probe branch was frequency quadrupled to 260~nm (4.77~eV) using two nonlinear crystals and focused onto a gold-coated sapphire in high vacuum ($<3.9\times10^{-9}$~torr) to generate pulsed photoelectron bunches. 
These electrons were accelerated to 26~kV in a DC field and focused with a solenoid onto the freestanding EuTe$_4$ flakes in transmission geometry. 
Diffracted electrons were collected by an aluminum-coated phosphor screen (P-46), whose luminescence was recorded by a commercial intensified charge-coupled device (iCCD, PI-MAX~II, Princeton Instruments) operating in the shutter mode. \textcolor{black}{The nominal momentum resolution, estimated from the full width half maximum of the raw electron beam, is $\sim$ 0.03 \AA$^{-1}$. Super-resolution information down to $\sim$ 1$\times10^{-5}$ \AA$^{-1}$ could be acquired with a regular Lorentzian fitting, as demonstrated in our previous UED experiments with the same instrument.\cite{Su2023}}
More details of the UED setup can be found elsewhere \cite{Freelon2023DesignInstrument}.
%MeV-UED experiments were carried out at the MeV-UED facility at SLAC National Accelerator Laboratory \cite{Weathersby2015Mega-electron-voltLaboratory}. The 800~nm (1.55~eV), 80~fs output from a commercial Ti:sapphire regenerative amplifier laser (Vitara and Legend Elite HE, Coherent Inc.) was split into the pump and probe arms. The pulse in the probe arm was frequency-tripled and was used to generate 4.2~MeV electron bunches via photoemission from a copper target. The photoelectrons were then accelerated in a radio-frequency photo-injector at a repetition rate of 360~Hz. The diffraction pattern was imaged by a phosphor screen (P-43) and recorded by an electron-multiplying charge-coupled device (EMCCD, Andor iXon Ultra 888). A circular through hole in the center of the phosphor screen allowed the passage of undiffracted electron beam to prevent camera saturation. The overall temporal resolution was approximately 300~fs.

\paragraph{Tr-ARPES experiment\\}

The 1030~nm (1.2~eV) output of either a Yb:KGW laser system (Pharos, Lightconversion) or a Yb-fiber laser system (Tangerine, Amplitude) was split into pump and probe branches, respectively. 
The pump branch was focused onto the cleaved surface of EuTe$_4$ to excite the sample \textcolor{black}{(s-polarization)}. 
The probe branch was first frequency-tripled to 344~nm (3.6~eV) and then focused onto either the hollow-core fiber filled with Xe gas or the Ar gas ejected by a gas-jet nozzle to generate the 9th harmonic at 114~nm (10.8~eV). 
The resulting extreme-ultraviolet (XUV) pulse was passed through a custom-built time-preserving grating monochromator (McPherson OP-XCT) to minimize pulse width broadening and enhance efficiency. 
After exiting the monochromator, the XUV pulse was focused onto the sample surface by a toroidal mirror. 
Photoelectrons were collected by a time-of-flight detector (Scienta ARTOF 10k). More details of the tr-ARPES setup can be found elsewhere \cite{Lee2020HighPulses,Sie2019Time-resolvedXUVResolution}.
\textcolor{black}{We note that although the data shown here were acquired in the s-polarization geometry, complementary p-polarization measurements yield no discernible differences in the observed dynamics.}

\section*{Data Availability}
\noindent The datasets generated and/or analyzed during the current study are available from the corresponding author on request.

\section*{Code Availability}
\noindent The codes used for the current study are available from the corresponding author on request.

\section*{Acknowledgments}
\noindent The authors thank Zhengyan Darius Shi, Tianchuang Luo, Batyr Ilyas, Bryan Fichera, Mingu Kang and Zhuquan Zhang for fruitful discussions. The work at MIT was supported by the US Department of Energy, Materials Science and Engineering Division, Office of Basic Energy Sciences (BES DMSE). N.L.W acknowledges support from National Natural Science Foundation of China (Grant No. 12488201). N.L.W and D.W are supported by National Key Research and Development Program of China (2024YFA1408700). H.J. acknowledges support by the National Research Foundation grant funded by the Korea government (MSIT) (Grant No. RS-2022-NR068223). The tr-XRD experiments were performed at the SSS-RSXS end station (Proposal No. 2024-1st-SSS-008) of the PAL-XFEL funded by the Korea government (MSIT).
The static XRD experiments were conducted at the Center for High-Energy X-ray Sciences (CHEXS), which is supported by the National Science Foundation (BIO, ENG and MPS Directorates) under award DMR-2342336. Use of the LCLS, SLAC National Accelerator Laboratory, is supported by the U.S. Department of Energy, Office of Science, Office of Basic Energy Sciences under Contract No.DE-AC02-76SF00515.

\section*{Author Contributions}
\noindent K.H.O., Y.S., H.N., G.K., H.-W.J.K., S.H., J.K., and H.J. performed the tr-XRD experiment. Y.S., B.Q.L., and A.Z. performed static XRD experiments. K.H.O., Y.S., and H.N. performed keV-UED experiments. Y.S., A.Z., and K.H.O. performed static ED experiments. Y.S., B.Q.L., A.Z., D.C., D.A., and M.M. performed tr-ARPES experiments. D.W. and Q.L. synthesized, characterized, and prepared the samples for the experiment. G.K., H.C., and H.J. maintained the beamline for the tr-XRD experiments. S.S. and J.P.C.R. maintained the beamline for the static XRD experiments. X.S., D.L., S.W., P.K., and X.C. maintained the beamline for the static ED experiments. B.Q.L., Y.S., D.C., D.A., and M.M. developed and maintained the tr-ARPES setup. K.H.O., Y.S., and H.N. analyzed the data. K.H.O., Y.S., and H.N. wrote the manuscript with critical input from B.J.K., N.L.W., H.J., N.G., and all other authors. The work was supervised by N.G.

\section*{Competing Interests}
\noindent The authors declare no competing interests.

\newpage

\begin{onehalfspace}

\newpage

\newpage

\newpage

\newpage

\end{onehalfspace}

%TC:endignore

\begin{thebibliography}{53}%
\makeatletter
\providecommand \@ifxundefined [1]{%
 \@ifx{#1\undefined}
}%
\providecommand \@ifnum [1]{%
 \ifnum #1\expandafter \@firstoftwo
 \else \expandafter \@secondoftwo
 \fi
}%
\providecommand \@ifx [1]{%
 \ifx #1\expandafter \@firstoftwo
 \else \expandafter \@secondoftwo
 \fi
}%
\providecommand \natexlab [1]{#1}%
\providecommand \enquote  [1]{``#1''}%
\providecommand \bibnamefont  [1]{#1}%
\providecommand \bibfnamefont [1]{#1}%
\providecommand \citenamefont [1]{#1}%
\providecommand \href@noop [0]{\@secondoftwo}%
\providecommand \href [0]{\begingroup \@sanitize@url \@href}%
\providecommand \@href[1]{\@@startlink{#1}\@@href}%
\providecommand \@@href[1]{\endgroup#1\@@endlink}%
\providecommand \@sanitize@url [0]{\catcode `\\12\catcode `\$12\catcode `\&12\catcode `\#12\catcode `\^12\catcode `\_12\catcode `\%12\relax}%
\providecommand \@@startlink[1]{}%
\providecommand \@@endlink[0]{}%
\providecommand \url  [0]{\begingroup\@sanitize@url \@url }%
\providecommand \@url [1]{\endgroup\@href {#1}{\urlprefix }}%
\providecommand \urlprefix  [0]{URL }%
\providecommand \Eprint [0]{\href }%
\providecommand \doibase [0]{https://doi.org/}%
\providecommand \selectlanguage [0]{\@gobble}%
\providecommand \bibinfo  [0]{\@secondoftwo}%
\providecommand \bibfield  [0]{\@secondoftwo}%
\providecommand \translation [1]{[#1]}%
\providecommand \BibitemOpen [0]{}%
\providecommand \bibitemStop [0]{}%
\providecommand \bibitemNoStop [0]{.\EOS\space}%
\providecommand \EOS [0]{\spacefactor3000\relax}%
\providecommand \BibitemShut  [1]{\csname bibitem#1\endcsname}%
\let\auto@bib@innerbib\@empty
%</preamble>
\bibitem [{\citenamefont {Cao}\ \emph {et~al.}(2018{\natexlab{a}})\citenamefont {Cao}, \citenamefont {Fatemi}, \citenamefont {Fang}, \citenamefont {Watanabe}, \citenamefont {Taniguchi}, \citenamefont {Kaxiras},\ and\ \citenamefont {Jarillo-Herrero}}]{Cao2018UnconventionalSuperlattices}%
  \BibitemOpen
  \bibfield  {author} {\bibinfo {author} {\bibfnamefont {Y.}~\bibnamefont {Cao}}, \bibinfo {author} {\bibfnamefont {V.}~\bibnamefont {Fatemi}}, \bibinfo {author} {\bibfnamefont {S.}~\bibnamefont {Fang}}, \bibinfo {author} {\bibfnamefont {K.}~\bibnamefont {Watanabe}}, \bibinfo {author} {\bibfnamefont {T.}~\bibnamefont {Taniguchi}}, \bibinfo {author} {\bibfnamefont {E.}~\bibnamefont {Kaxiras}},\ and\ \bibinfo {author} {\bibfnamefont {P.}~\bibnamefont {Jarillo-Herrero}},\ }\bibfield  {title} {\bibinfo {title} {{Unconventional superconductivity in magic-angle graphene superlattices}},\ }\href {https://doi.org/10.1038/nature26160} {\bibfield  {journal} {\bibinfo  {journal} {Nature}\ }\textbf {\bibinfo {volume} {556}},\ \bibinfo {pages} {43} (\bibinfo {year} {2018}{\natexlab{a}})}\BibitemShut {NoStop}%
\bibitem [{\citenamefont {Cao}\ \emph {et~al.}(2018{\natexlab{b}})\citenamefont {Cao}, \citenamefont {Fatemi}, \citenamefont {Demir}, \citenamefont {Fang}, \citenamefont {Tomarken}, \citenamefont {Luo}, \citenamefont {Sanchez-Yamagishi}, \citenamefont {Watanabe}, \citenamefont {Taniguchi}, \citenamefont {Kaxiras}, \citenamefont {Ashoori},\ and\ \citenamefont {Jarillo-Herrero}}]{Cao2018CorrelatedSuperlattices}%
  \BibitemOpen
  \bibfield  {author} {\bibinfo {author} {\bibfnamefont {Y.}~\bibnamefont {Cao}}, \bibinfo {author} {\bibfnamefont {V.}~\bibnamefont {Fatemi}}, \bibinfo {author} {\bibfnamefont {A.}~\bibnamefont {Demir}}, \bibinfo {author} {\bibfnamefont {S.}~\bibnamefont {Fang}}, \bibinfo {author} {\bibfnamefont {S.~L.}\ \bibnamefont {Tomarken}}, \bibinfo {author} {\bibfnamefont {J.~Y.}\ \bibnamefont {Luo}}, \bibinfo {author} {\bibfnamefont {J.~D.}\ \bibnamefont {Sanchez-Yamagishi}}, \bibinfo {author} {\bibfnamefont {K.}~\bibnamefont {Watanabe}}, \bibinfo {author} {\bibfnamefont {T.}~\bibnamefont {Taniguchi}}, \bibinfo {author} {\bibfnamefont {E.}~\bibnamefont {Kaxiras}}, \bibinfo {author} {\bibfnamefont {R.~C.}\ \bibnamefont {Ashoori}},\ and\ \bibinfo {author} {\bibfnamefont {P.}~\bibnamefont {Jarillo-Herrero}},\ }\bibfield  {title} {\bibinfo {title} {{Correlated insulator behaviour at half-filling in magic-angle graphene superlattices}},\ }\href {https://doi.org/10.1038/nature26154} {\bibfield  {journal} {\bibinfo
  {journal} {Nature}\ }\textbf {\bibinfo {volume} {556}},\ \bibinfo {pages} {80} (\bibinfo {year} {2018}{\natexlab{b}})}\BibitemShut {NoStop}%
\bibitem [{\citenamefont {Mak}\ and\ \citenamefont {Shan}(2022)}]{Mak2022SemiconductorMaterials}%
  \BibitemOpen
  \bibfield  {author} {\bibinfo {author} {\bibfnamefont {K.~F.}\ \bibnamefont {Mak}}\ and\ \bibinfo {author} {\bibfnamefont {J.}~\bibnamefont {Shan}},\ }\bibfield  {title} {\bibinfo {title} {{Semiconductor moir{\'{e}} materials}},\ }\href {https://doi.org/10.1038/s41565-022-01165-6} {\bibfield  {journal} {\bibinfo  {journal} {Nature Nanotechnology}\ }\textbf {\bibinfo {volume} {17}},\ \bibinfo {pages} {686} (\bibinfo {year} {2022})}\BibitemShut {NoStop}%
\bibitem [{\citenamefont {Lv}\ \emph {et~al.}(2026)\citenamefont {Lv}, \citenamefont {Su}, \citenamefont {Zong}, \citenamefont {Liu}, \citenamefont {Wu}, \citenamefont {Yuan}, \citenamefont {Nie}, \citenamefont {Li}, \citenamefont {Sarker}, \citenamefont {Meng}, \citenamefont {Ruff}, \citenamefont {Wang},\ and\ \citenamefont {Gedik}}]{Lv2024LargeWaves}%
  \BibitemOpen
  \bibfield  {author} {\bibinfo {author} {\bibfnamefont {B.~Q.}\ \bibnamefont {Lv}}, \bibinfo {author} {\bibfnamefont {Y.}~\bibnamefont {Su}}, \bibinfo {author} {\bibfnamefont {A.}~\bibnamefont {Zong}}, \bibinfo {author} {\bibfnamefont {Q.}~\bibnamefont {Liu}}, \bibinfo {author} {\bibfnamefont {D.}~\bibnamefont {Wu}}, \bibinfo {author} {\bibfnamefont {N.~F.~Q.}\ \bibnamefont {Yuan}}, \bibinfo {author} {\bibfnamefont {Z.}~\bibnamefont {Nie}}, \bibinfo {author} {\bibfnamefont {J.}~\bibnamefont {Li}}, \bibinfo {author} {\bibfnamefont {S.}~\bibnamefont {Sarker}}, \bibinfo {author} {\bibfnamefont {S.}~\bibnamefont {Meng}}, \bibinfo {author} {\bibfnamefont {J.~P.~C.}\ \bibnamefont {Ruff}}, \bibinfo {author} {\bibfnamefont {N.~L.}\ \bibnamefont {Wang}},\ and\ \bibinfo {author} {\bibfnamefont {N.}~\bibnamefont {Gedik}},\ }\bibfield  {title} {\bibinfo {title} {{Large moir{\'{e}} superstructure of stacked incommensurate charge density waves}},\ }\href {https://doi.org/10.1038/s41563-025-02360-1} {\bibfield  {journal}
  {\bibinfo  {journal} {Nature Materials}\ }\textbf {\bibinfo {volume} {25}},\ \bibinfo {pages} {420} (\bibinfo {year} {2026})},\ \Eprint {https://arxiv.org/abs/2501.09715} {2501.09715} \BibitemShut {NoStop}%
\bibitem [{\citenamefont {Emery}\ and\ \citenamefont {Mukamel}(1979)}]{Emery1979Lockingsub3/sub}%
  \BibitemOpen
  \bibfield  {author} {\bibinfo {author} {\bibfnamefont {V.~J.}\ \bibnamefont {Emery}}\ and\ \bibinfo {author} {\bibfnamefont {D.}~\bibnamefont {Mukamel}},\ }\bibfield  {title} {\bibinfo {title} {{Locking of the two charge-density waves in NbSe$_3$}},\ }\href {https://doi.org/10.1088/0022-3719/12/17/008} {\bibfield  {journal} {\bibinfo  {journal} {Journal of Physics C : Solid State Physics}\ }\textbf {\bibinfo {volume} {12}},\ \bibinfo {pages} {L677} (\bibinfo {year} {1979})}\BibitemShut {NoStop}%
\bibitem [{\citenamefont {Lee}\ and\ \citenamefont {Rice}(1979)}]{Lee1979ElectricWaves}%
  \BibitemOpen
  \bibfield  {author} {\bibinfo {author} {\bibfnamefont {P.~A.}\ \bibnamefont {Lee}}\ and\ \bibinfo {author} {\bibfnamefont {T.~M.}\ \bibnamefont {Rice}},\ }\bibfield  {title} {\bibinfo {title} {{Electric field depinning of charge density waves}},\ }\href {https://doi.org/10.1103/PhysRevB.19.3970} {\bibfield  {journal} {\bibinfo  {journal} {Physical Review B}\ }\textbf {\bibinfo {volume} {19}},\ \bibinfo {pages} {3970} (\bibinfo {year} {1979})}\BibitemShut {NoStop}%
\bibitem [{\citenamefont {Bruinsma}\ and\ \citenamefont {Trullinger}(1980)}]{Bruinsma1980Phase-locked/math}%
  \BibitemOpen
  \bibfield  {author} {\bibinfo {author} {\bibfnamefont {R.}~\bibnamefont {Bruinsma}}\ and\ \bibinfo {author} {\bibfnamefont {S.~E.}\ \bibnamefont {Trullinger}},\ }\bibfield  {title} {\bibinfo {title} {{Phase-locked charge-density waves in NbSe$_3$}},\ }\href {https://doi.org/10.1103/PhysRevB.22.4543} {\bibfield  {journal} {\bibinfo  {journal} {Physical Review B}\ }\textbf {\bibinfo {volume} {22}},\ \bibinfo {pages} {4543} (\bibinfo {year} {1980})}\BibitemShut {NoStop}%
\bibitem [{\citenamefont {Ayari}\ \emph {et~al.}(2004)\citenamefont {Ayari}, \citenamefont {Danneau}, \citenamefont {Requardt}, \citenamefont {Ortega}, \citenamefont {Lorenzo}, \citenamefont {Monceau}, \citenamefont {Currat}, \citenamefont {Brazovskii},\ and\ \citenamefont {Gr{\"{u}}bel}}]{Ayari2004Sliding-inducedNbSe3}%
  \BibitemOpen
  \bibfield  {author} {\bibinfo {author} {\bibfnamefont {A.}~\bibnamefont {Ayari}}, \bibinfo {author} {\bibfnamefont {R.}~\bibnamefont {Danneau}}, \bibinfo {author} {\bibfnamefont {H.}~\bibnamefont {Requardt}}, \bibinfo {author} {\bibfnamefont {L.}~\bibnamefont {Ortega}}, \bibinfo {author} {\bibfnamefont {J.~E.}\ \bibnamefont {Lorenzo}}, \bibinfo {author} {\bibfnamefont {P.}~\bibnamefont {Monceau}}, \bibinfo {author} {\bibfnamefont {R.}~\bibnamefont {Currat}}, \bibinfo {author} {\bibfnamefont {S.}~\bibnamefont {Brazovskii}},\ and\ \bibinfo {author} {\bibfnamefont {G.}~\bibnamefont {Gr{\"{u}}bel}},\ }\bibfield  {title} {\bibinfo {title} {{Sliding-induced decoupling and charge transfer between the coexisting Q$_1$ and Q$_2$ charge density waves in NbSe$_3$}},\ }\href {https://doi.org/10.1103/PhysRevLett.93.106404} {\bibfield  {journal} {\bibinfo  {journal} {Physical Review Letters}\ }\textbf {\bibinfo {volume} {93}},\ \bibinfo {pages} {106404} (\bibinfo {year} {2004})}\BibitemShut {NoStop}%
\bibitem [{\citenamefont {Kivelson}\ \emph {et~al.}(1998)\citenamefont {Kivelson}, \citenamefont {Fradkin},\ and\ \citenamefont {Emery}}]{Kivelson1998ElectronicInsulator}%
  \BibitemOpen
  \bibfield  {author} {\bibinfo {author} {\bibfnamefont {S.~A.}\ \bibnamefont {Kivelson}}, \bibinfo {author} {\bibfnamefont {E.}~\bibnamefont {Fradkin}},\ and\ \bibinfo {author} {\bibfnamefont {V.~J.}\ \bibnamefont {Emery}},\ }\bibfield  {title} {\bibinfo {title} {{Electronic liquid-crystal phases of a doped Mott insulator}},\ }\href {https://doi.org/10.1038/31177} {\bibfield  {journal} {\bibinfo  {journal} {Nature}\ }\textbf {\bibinfo {volume} {393}},\ \bibinfo {pages} {550} (\bibinfo {year} {1998})}\BibitemShut {NoStop}%
\bibitem [{\citenamefont {Fradkin}\ \emph {et~al.}(2010)\citenamefont {Fradkin}, \citenamefont {Kivelson}, \citenamefont {Lawler}, \citenamefont {Eisenstein},\ and\ \citenamefont {Mackenzie}}]{Fradkin2010NematicPhysics}%
  \BibitemOpen
  \bibfield  {author} {\bibinfo {author} {\bibfnamefont {E.}~\bibnamefont {Fradkin}}, \bibinfo {author} {\bibfnamefont {S.~A.}\ \bibnamefont {Kivelson}}, \bibinfo {author} {\bibfnamefont {M.~J.}\ \bibnamefont {Lawler}}, \bibinfo {author} {\bibfnamefont {J.~P.}\ \bibnamefont {Eisenstein}},\ and\ \bibinfo {author} {\bibfnamefont {A.~P.}\ \bibnamefont {Mackenzie}},\ }\bibfield  {title} {\bibinfo {title} {{Nematic Fermi Fluids in Condensed Matter Physics}},\ }\href {https://doi.org/10.1146/annurev-conmatphys-070909-103925} {\bibfield  {journal} {\bibinfo  {journal} {Annual Review of Condensed Matter Physics}\ }\textbf {\bibinfo {volume} {1}},\ \bibinfo {pages} {153} (\bibinfo {year} {2010})}\BibitemShut {NoStop}%
\bibitem [{\citenamefont {Lv}\ \emph {et~al.}(2022)\citenamefont {Lv}, \citenamefont {Zong}, \citenamefont {Wu}, \citenamefont {Rozhkov}, \citenamefont {Fine}, \citenamefont {Chen}, \citenamefont {Hashimoto}, \citenamefont {Lu}, \citenamefont {Li}, \citenamefont {Huang}, \citenamefont {Ruff}, \citenamefont {Walko}, \citenamefont {Chen}, \citenamefont {Hwang}, \citenamefont {Su}, \citenamefont {Shen}, \citenamefont {Wang}, \citenamefont {Han}, \citenamefont {Po}, \citenamefont {Wang}, \citenamefont {Jarillo-Herrero}, \citenamefont {Wang}, \citenamefont {Zhou}, \citenamefont {Sun}, \citenamefont {Wen}, \citenamefont {Shen}, \citenamefont {Wang},\ and\ \citenamefont {Gedik}}]{Lv2022UnconventionalWave}%
  \BibitemOpen
  \bibfield  {author} {\bibinfo {author} {\bibfnamefont {B.}~\bibnamefont {Lv}}, \bibinfo {author} {\bibfnamefont {A.}~\bibnamefont {Zong}}, \bibinfo {author} {\bibfnamefont {D.}~\bibnamefont {Wu}}, \bibinfo {author} {\bibfnamefont {A.}~\bibnamefont {Rozhkov}}, \bibinfo {author} {\bibfnamefont {B.~V.}\ \bibnamefont {Fine}}, \bibinfo {author} {\bibfnamefont {S.-D.}\ \bibnamefont {Chen}}, \bibinfo {author} {\bibfnamefont {M.}~\bibnamefont {Hashimoto}}, \bibinfo {author} {\bibfnamefont {D.-H.}\ \bibnamefont {Lu}}, \bibinfo {author} {\bibfnamefont {M.}~\bibnamefont {Li}}, \bibinfo {author} {\bibfnamefont {Y.-B.}\ \bibnamefont {Huang}}, \bibinfo {author} {\bibfnamefont {J.~P.}\ \bibnamefont {Ruff}}, \bibinfo {author} {\bibfnamefont {D.~A.}\ \bibnamefont {Walko}}, \bibinfo {author} {\bibfnamefont {Z.}~\bibnamefont {Chen}}, \bibinfo {author} {\bibfnamefont {I.}~\bibnamefont {Hwang}}, \bibinfo {author} {\bibfnamefont {Y.}~\bibnamefont {Su}}, \bibinfo {author} {\bibfnamefont {X.}~\bibnamefont {Shen}}, \bibinfo
  {author} {\bibfnamefont {X.}~\bibnamefont {Wang}}, \bibinfo {author} {\bibfnamefont {F.}~\bibnamefont {Han}}, \bibinfo {author} {\bibfnamefont {H.~C.}\ \bibnamefont {Po}}, \bibinfo {author} {\bibfnamefont {Y.}~\bibnamefont {Wang}}, \bibinfo {author} {\bibfnamefont {P.}~\bibnamefont {Jarillo-Herrero}}, \bibinfo {author} {\bibfnamefont {X.}~\bibnamefont {Wang}}, \bibinfo {author} {\bibfnamefont {H.}~\bibnamefont {Zhou}}, \bibinfo {author} {\bibfnamefont {C.-J.}\ \bibnamefont {Sun}}, \bibinfo {author} {\bibfnamefont {H.}~\bibnamefont {Wen}}, \bibinfo {author} {\bibfnamefont {Z.-X.}\ \bibnamefont {Shen}}, \bibinfo {author} {\bibfnamefont {N.}~\bibnamefont {Wang}},\ and\ \bibinfo {author} {\bibfnamefont {N.}~\bibnamefont {Gedik}},\ }\bibfield  {title} {\bibinfo {title} {{Unconventional Hysteretic Transition in a Charge Density Wave}},\ }\href {https://doi.org/10.1103/PhysRevLett.128.036401} {\bibfield  {journal} {\bibinfo  {journal} {Physical Review Letters}\ }\textbf {\bibinfo {volume} {128}},\ \bibinfo {pages}
  {036401} (\bibinfo {year} {2022})}\BibitemShut {NoStop}%
\bibitem [{\citenamefont {Wu}\ \emph {et~al.}(2019)\citenamefont {Wu}, \citenamefont {Liu}, \citenamefont {Chen}, \citenamefont {Zhong}, \citenamefont {Su}, \citenamefont {Shi}, \citenamefont {Tong}, \citenamefont {Xu}, \citenamefont {Gao},\ and\ \citenamefont {Wang}}]{Wu2019LayeredSheets}%
  \BibitemOpen
  \bibfield  {author} {\bibinfo {author} {\bibfnamefont {D.}~\bibnamefont {Wu}}, \bibinfo {author} {\bibfnamefont {Q.~M.}\ \bibnamefont {Liu}}, \bibinfo {author} {\bibfnamefont {S.~L.}\ \bibnamefont {Chen}}, \bibinfo {author} {\bibfnamefont {G.~Y.}\ \bibnamefont {Zhong}}, \bibinfo {author} {\bibfnamefont {J.}~\bibnamefont {Su}}, \bibinfo {author} {\bibfnamefont {L.~Y.}\ \bibnamefont {Shi}}, \bibinfo {author} {\bibfnamefont {L.}~\bibnamefont {Tong}}, \bibinfo {author} {\bibfnamefont {G.}~\bibnamefont {Xu}}, \bibinfo {author} {\bibfnamefont {P.}~\bibnamefont {Gao}},\ and\ \bibinfo {author} {\bibfnamefont {N.~L.}\ \bibnamefont {Wang}},\ }\bibfield  {title} {\bibinfo {title} {{Layered semiconductor EuTe$_4$ with charge density wave order in square tellurium sheets}},\ }\href {https://doi.org/10.1103/PhysRevMaterials.3.024002} {\bibfield  {journal} {\bibinfo  {journal} {Physical Review Materials}\ }\textbf {\bibinfo {volume} {3}},\ \bibinfo {pages} {024002} (\bibinfo {year} {2019})}\BibitemShut {NoStop}%
\bibitem [{\citenamefont {Liu}\ \emph {et~al.}(2024)\citenamefont {Liu}, \citenamefont {Wu}, \citenamefont {Wu}, \citenamefont {Han}, \citenamefont {Peng}, \citenamefont {Yuan}, \citenamefont {Cheng}, \citenamefont {Li}, \citenamefont {Hu}, \citenamefont {Yue}, \citenamefont {Xu}, \citenamefont {Ding}, \citenamefont {Lu}, \citenamefont {Li}, \citenamefont {Zhang}, \citenamefont {Lv}, \citenamefont {Zong}, \citenamefont {Su}, \citenamefont {Gedik}, \citenamefont {Yin}, \citenamefont {Dong},\ and\ \citenamefont {Wang}}]{Liu2024}%
  \BibitemOpen
  \bibfield  {author} {\bibinfo {author} {\bibfnamefont {Q.}~\bibnamefont {Liu}}, \bibinfo {author} {\bibfnamefont {D.}~\bibnamefont {Wu}}, \bibinfo {author} {\bibfnamefont {T.}~\bibnamefont {Wu}}, \bibinfo {author} {\bibfnamefont {S.}~\bibnamefont {Han}}, \bibinfo {author} {\bibfnamefont {Y.}~\bibnamefont {Peng}}, \bibinfo {author} {\bibfnamefont {Z.}~\bibnamefont {Yuan}}, \bibinfo {author} {\bibfnamefont {Y.}~\bibnamefont {Cheng}}, \bibinfo {author} {\bibfnamefont {B.}~\bibnamefont {Li}}, \bibinfo {author} {\bibfnamefont {T.}~\bibnamefont {Hu}}, \bibinfo {author} {\bibfnamefont {L.}~\bibnamefont {Yue}}, \bibinfo {author} {\bibfnamefont {S.}~\bibnamefont {Xu}}, \bibinfo {author} {\bibfnamefont {R.}~\bibnamefont {Ding}}, \bibinfo {author} {\bibfnamefont {M.}~\bibnamefont {Lu}}, \bibinfo {author} {\bibfnamefont {R.}~\bibnamefont {Li}}, \bibinfo {author} {\bibfnamefont {S.}~\bibnamefont {Zhang}}, \bibinfo {author} {\bibfnamefont {B.}~\bibnamefont {Lv}}, \bibinfo {author} {\bibfnamefont {A.}~\bibnamefont
  {Zong}}, \bibinfo {author} {\bibfnamefont {Y.}~\bibnamefont {Su}}, \bibinfo {author} {\bibfnamefont {N.}~\bibnamefont {Gedik}}, \bibinfo {author} {\bibfnamefont {Z.}~\bibnamefont {Yin}}, \bibinfo {author} {\bibfnamefont {T.}~\bibnamefont {Dong}},\ and\ \bibinfo {author} {\bibfnamefont {N.}~\bibnamefont {Wang}},\ }\bibfield  {title} {\bibinfo {title} {{Room-temperature non-volatile optical manipulation of polar order in a charge density wave}},\ }\href {https://doi.org/10.1038/s41467-024-53323-0} {\bibfield  {journal} {\bibinfo  {journal} {Nature Communications}\ }\textbf {\bibinfo {volume} {15}},\ \bibinfo {pages} {8937} (\bibinfo {year} {2024})}\BibitemShut {NoStop}%
\bibitem [{\citenamefont {Venturini}\ \emph {et~al.}(2024)\citenamefont {Venturini}, \citenamefont {Rupnik}, \citenamefont {Gasperlin}, \citenamefont {Lipic}, \citenamefont {Sutar}, \citenamefont {Vaskivskyi}, \citenamefont {Scepanovic}, \citenamefont {Grabnar}, \citenamefont {Golez},\ and\ \citenamefont {Mihailovic}}]{Venturini2024}%
  \BibitemOpen
  \bibfield  {author} {\bibinfo {author} {\bibfnamefont {R.}~\bibnamefont {Venturini}}, \bibinfo {author} {\bibfnamefont {M.}~\bibnamefont {Rupnik}}, \bibinfo {author} {\bibfnamefont {J.}~\bibnamefont {Gasperlin}}, \bibinfo {author} {\bibfnamefont {J.}~\bibnamefont {Lipic}}, \bibinfo {author} {\bibfnamefont {P.}~\bibnamefont {Sutar}}, \bibinfo {author} {\bibfnamefont {Y.}~\bibnamefont {Vaskivskyi}}, \bibinfo {author} {\bibfnamefont {F.}~\bibnamefont {Scepanovic}}, \bibinfo {author} {\bibfnamefont {D.}~\bibnamefont {Grabnar}}, \bibinfo {author} {\bibfnamefont {D.}~\bibnamefont {Golez}},\ and\ \bibinfo {author} {\bibfnamefont {D.}~\bibnamefont {Mihailovic}},\ }\bibfield  {title} {\bibinfo {title} {Electrically driven non-volatile resistance switching between charge density wave states at room temperature},\ }\href {https://arxiv.org/abs/2412.13094} {\bibfield  {journal} {\bibinfo  {journal} {arXiv:2412.13094}\ } (\bibinfo {year} {2024})}\BibitemShut {NoStop}%
\bibitem [{\citenamefont {Duncan}\ \emph {et~al.}(2025)\citenamefont {Duncan}, \citenamefont {Johnson}, \citenamefont {Maity}, \citenamefont {Rubio}, \citenamefont {Gordon}, \citenamefont {Bartnik}, \citenamefont {Kaemingk}, \citenamefont {Li}, \citenamefont {Andorf}, \citenamefont {Pennington}, \citenamefont {Bazarov}, \citenamefont {Tate}, \citenamefont {Muller}, \citenamefont {Thom-Levy}, \citenamefont {Gruner}, \citenamefont {Lindenberg}, \citenamefont {Maxson},\ and\ \citenamefont {Liu}}]{Duncan2025PhotoinducedTwist}%
  \BibitemOpen
  \bibfield  {author} {\bibinfo {author} {\bibfnamefont {C.~J.~R.}\ \bibnamefont {Duncan}}, \bibinfo {author} {\bibfnamefont {A.~C.}\ \bibnamefont {Johnson}}, \bibinfo {author} {\bibfnamefont {I.}~\bibnamefont {Maity}}, \bibinfo {author} {\bibfnamefont {A.}~\bibnamefont {Rubio}}, \bibinfo {author} {\bibfnamefont {M.}~\bibnamefont {Gordon}}, \bibinfo {author} {\bibfnamefont {A.~C.}\ \bibnamefont {Bartnik}}, \bibinfo {author} {\bibfnamefont {M.}~\bibnamefont {Kaemingk}}, \bibinfo {author} {\bibfnamefont {W.~H.}\ \bibnamefont {Li}}, \bibinfo {author} {\bibfnamefont {M.~B.}\ \bibnamefont {Andorf}}, \bibinfo {author} {\bibfnamefont {C.~A.}\ \bibnamefont {Pennington}}, \bibinfo {author} {\bibfnamefont {I.~V.}\ \bibnamefont {Bazarov}}, \bibinfo {author} {\bibfnamefont {M.~W.}\ \bibnamefont {Tate}}, \bibinfo {author} {\bibfnamefont {D.~A.}\ \bibnamefont {Muller}}, \bibinfo {author} {\bibfnamefont {J.}~\bibnamefont {Thom-Levy}}, \bibinfo {author} {\bibfnamefont {S.~M.}\ \bibnamefont {Gruner}}, \bibinfo {author}
  {\bibfnamefont {A.~M.}\ \bibnamefont {Lindenberg}}, \bibinfo {author} {\bibfnamefont {J.~M.}\ \bibnamefont {Maxson}},\ and\ \bibinfo {author} {\bibfnamefont {F.}~\bibnamefont {Liu}},\ }\bibfield  {title} {\bibinfo {title} {Photoinduced twist and untwist of moir{\'e} superlattices},\ }\href {https://doi.org/10.1038/s41586-025-09707-3} {\bibfield  {journal} {\bibinfo  {journal} {Nature}\ }\textbf {\bibinfo {volume} {647}},\ \bibinfo {pages} {619} (\bibinfo {year} {2025})}\BibitemShut {NoStop}%
\bibitem [{\citenamefont {Zong}\ \emph {et~al.}(2018)\citenamefont {Zong}, \citenamefont {Shen}, \citenamefont {Kogar}, \citenamefont {Ye}, \citenamefont {Marks}, \citenamefont {Chowdhury}, \citenamefont {Rohwer}, \citenamefont {Freelon}, \citenamefont {Weathersby}, \citenamefont {Li}, \citenamefont {Yang}, \citenamefont {Checkelsky}, \citenamefont {Wang},\ and\ \citenamefont {Gedik}}]{Zong2018UltrafastWave}%
  \BibitemOpen
  \bibfield  {author} {\bibinfo {author} {\bibfnamefont {A.}~\bibnamefont {Zong}}, \bibinfo {author} {\bibfnamefont {X.}~\bibnamefont {Shen}}, \bibinfo {author} {\bibfnamefont {A.}~\bibnamefont {Kogar}}, \bibinfo {author} {\bibfnamefont {L.}~\bibnamefont {Ye}}, \bibinfo {author} {\bibfnamefont {C.}~\bibnamefont {Marks}}, \bibinfo {author} {\bibfnamefont {D.}~\bibnamefont {Chowdhury}}, \bibinfo {author} {\bibfnamefont {T.}~\bibnamefont {Rohwer}}, \bibinfo {author} {\bibfnamefont {B.}~\bibnamefont {Freelon}}, \bibinfo {author} {\bibfnamefont {S.}~\bibnamefont {Weathersby}}, \bibinfo {author} {\bibfnamefont {R.}~\bibnamefont {Li}}, \bibinfo {author} {\bibfnamefont {J.}~\bibnamefont {Yang}}, \bibinfo {author} {\bibfnamefont {J.}~\bibnamefont {Checkelsky}}, \bibinfo {author} {\bibfnamefont {X.}~\bibnamefont {Wang}},\ and\ \bibinfo {author} {\bibfnamefont {N.}~\bibnamefont {Gedik}},\ }\bibfield  {title} {\bibinfo {title} {{Ultrafast manipulation of mirror domain walls in a charge density wave}},\ }\href
  {https://doi.org/10.1126/sciadv.aau5501} {\bibfield  {journal} {\bibinfo  {journal} {Science Advances}\ }\textbf {\bibinfo {volume} {4}},\ \bibinfo {pages} {eaau5501} (\bibinfo {year} {2018})}\BibitemShut {NoStop}%
\bibitem [{\citenamefont {Geremew}\ \emph {et~al.}(2019)\citenamefont {Geremew}, \citenamefont {Rumyantsev}, \citenamefont {Kargar}, \citenamefont {Debnath}, \citenamefont {Nosek}, \citenamefont {Bloodgood}, \citenamefont {Bockrath}, \citenamefont {Salguero}, \citenamefont {Lake},\ and\ \citenamefont {Balandin}}]{Geremew2019Bias-VoltageDevices}%
  \BibitemOpen
  \bibfield  {author} {\bibinfo {author} {\bibfnamefont {A.~K.}\ \bibnamefont {Geremew}}, \bibinfo {author} {\bibfnamefont {S.}~\bibnamefont {Rumyantsev}}, \bibinfo {author} {\bibfnamefont {F.}~\bibnamefont {Kargar}}, \bibinfo {author} {\bibfnamefont {B.}~\bibnamefont {Debnath}}, \bibinfo {author} {\bibfnamefont {A.}~\bibnamefont {Nosek}}, \bibinfo {author} {\bibfnamefont {M.~A.}\ \bibnamefont {Bloodgood}}, \bibinfo {author} {\bibfnamefont {M.}~\bibnamefont {Bockrath}}, \bibinfo {author} {\bibfnamefont {T.~T.}\ \bibnamefont {Salguero}}, \bibinfo {author} {\bibfnamefont {R.~K.}\ \bibnamefont {Lake}},\ and\ \bibinfo {author} {\bibfnamefont {A.~A.}\ \bibnamefont {Balandin}},\ }\bibfield  {title} {\bibinfo {title} {{Bias-Voltage Driven Switching of the Charge-Density-Wave and Normal Metallic Phases in 1T-TaS$_2$ Thin-Film Devices}},\ }\href {https://doi.org/10.1021/acsnano.9b02870} {\bibfield  {journal} {\bibinfo  {journal} {ACS Nano}\ }\textbf {\bibinfo {volume} {13}},\ \bibinfo {pages} {7231} (\bibinfo {year}
  {2019})}\BibitemShut {NoStop}%
\bibitem [{\citenamefont {Zong}\ \emph {et~al.}(2019)\citenamefont {Zong}, \citenamefont {Kogar}, \citenamefont {Bie}, \citenamefont {Rohwer}, \citenamefont {Lee}, \citenamefont {Baldini}, \citenamefont {Erge{\c{c}}en}, \citenamefont {Yilmaz}, \citenamefont {Freelon}, \citenamefont {Sie}, \citenamefont {Zhou}, \citenamefont {Straquadine}, \citenamefont {Walmsley}, \citenamefont {Dolgirev}, \citenamefont {Rozhkov}, \citenamefont {Fisher}, \citenamefont {Jarillo-Herrero}, \citenamefont {Fine},\ and\ \citenamefont {Gedik}}]{Zong2019EvidenceTransition}%
  \BibitemOpen
  \bibfield  {author} {\bibinfo {author} {\bibfnamefont {A.}~\bibnamefont {Zong}}, \bibinfo {author} {\bibfnamefont {A.}~\bibnamefont {Kogar}}, \bibinfo {author} {\bibfnamefont {Y.~Q.}\ \bibnamefont {Bie}}, \bibinfo {author} {\bibfnamefont {T.}~\bibnamefont {Rohwer}}, \bibinfo {author} {\bibfnamefont {C.}~\bibnamefont {Lee}}, \bibinfo {author} {\bibfnamefont {E.}~\bibnamefont {Baldini}}, \bibinfo {author} {\bibfnamefont {E.}~\bibnamefont {Erge{\c{c}}en}}, \bibinfo {author} {\bibfnamefont {M.~B.}\ \bibnamefont {Yilmaz}}, \bibinfo {author} {\bibfnamefont {B.}~\bibnamefont {Freelon}}, \bibinfo {author} {\bibfnamefont {E.~J.}\ \bibnamefont {Sie}}, \bibinfo {author} {\bibfnamefont {H.}~\bibnamefont {Zhou}}, \bibinfo {author} {\bibfnamefont {J.}~\bibnamefont {Straquadine}}, \bibinfo {author} {\bibfnamefont {P.}~\bibnamefont {Walmsley}}, \bibinfo {author} {\bibfnamefont {P.~E.}\ \bibnamefont {Dolgirev}}, \bibinfo {author} {\bibfnamefont {A.~V.}\ \bibnamefont {Rozhkov}}, \bibinfo {author} {\bibfnamefont {I.~R.}\
  \bibnamefont {Fisher}}, \bibinfo {author} {\bibfnamefont {P.}~\bibnamefont {Jarillo-Herrero}}, \bibinfo {author} {\bibfnamefont {B.~V.}\ \bibnamefont {Fine}},\ and\ \bibinfo {author} {\bibfnamefont {N.}~\bibnamefont {Gedik}},\ }\bibfield  {title} {\bibinfo {title} {{Evidence for topological defects in a photoinduced phase transition}},\ }\href {https://doi.org/10.1038/s41567-018-0311-9} {\bibfield  {journal} {\bibinfo  {journal} {Nature Physics}\ }\textbf {\bibinfo {volume} {15}},\ \bibinfo {pages} {27} (\bibinfo {year} {2019})}\BibitemShut {NoStop}%
\bibitem [{\citenamefont {Kogar}\ \emph {et~al.}(2020)\citenamefont {Kogar}, \citenamefont {Zong}, \citenamefont {Dolgirev}, \citenamefont {Shen}, \citenamefont {Straquadine}, \citenamefont {Bie}, \citenamefont {Wang}, \citenamefont {Rohwer}, \citenamefont {Tung}, \citenamefont {Yang}, \citenamefont {Li}, \citenamefont {Yang}, \citenamefont {Weathersby}, \citenamefont {Park}, \citenamefont {Kozina}, \citenamefont {Sie}, \citenamefont {Wen}, \citenamefont {Jarillo-Herrero}, \citenamefont {Fisher}, \citenamefont {Wang},\ and\ \citenamefont {Gedik}}]{Kogar2020Light-inducedLaTe3}%
  \BibitemOpen
  \bibfield  {author} {\bibinfo {author} {\bibfnamefont {A.}~\bibnamefont {Kogar}}, \bibinfo {author} {\bibfnamefont {A.}~\bibnamefont {Zong}}, \bibinfo {author} {\bibfnamefont {P.~E.}\ \bibnamefont {Dolgirev}}, \bibinfo {author} {\bibfnamefont {X.}~\bibnamefont {Shen}}, \bibinfo {author} {\bibfnamefont {J.}~\bibnamefont {Straquadine}}, \bibinfo {author} {\bibfnamefont {Y.~Q.}\ \bibnamefont {Bie}}, \bibinfo {author} {\bibfnamefont {X.}~\bibnamefont {Wang}}, \bibinfo {author} {\bibfnamefont {T.}~\bibnamefont {Rohwer}}, \bibinfo {author} {\bibfnamefont {I.~C.}\ \bibnamefont {Tung}}, \bibinfo {author} {\bibfnamefont {Y.}~\bibnamefont {Yang}}, \bibinfo {author} {\bibfnamefont {R.}~\bibnamefont {Li}}, \bibinfo {author} {\bibfnamefont {J.}~\bibnamefont {Yang}}, \bibinfo {author} {\bibfnamefont {S.}~\bibnamefont {Weathersby}}, \bibinfo {author} {\bibfnamefont {S.}~\bibnamefont {Park}}, \bibinfo {author} {\bibfnamefont {M.~E.}\ \bibnamefont {Kozina}}, \bibinfo {author} {\bibfnamefont {E.~J.}\ \bibnamefont {Sie}},
  \bibinfo {author} {\bibfnamefont {H.}~\bibnamefont {Wen}}, \bibinfo {author} {\bibfnamefont {P.}~\bibnamefont {Jarillo-Herrero}}, \bibinfo {author} {\bibfnamefont {I.~R.}\ \bibnamefont {Fisher}}, \bibinfo {author} {\bibfnamefont {X.}~\bibnamefont {Wang}},\ and\ \bibinfo {author} {\bibfnamefont {N.}~\bibnamefont {Gedik}},\ }\bibfield  {title} {\bibinfo {title} {{Light-induced charge density wave in LaTe$_3$}},\ }\href {https://doi.org/10.1038/s41567-019-0705-3} {\bibfield  {journal} {\bibinfo  {journal} {Nature Physics}\ }\textbf {\bibinfo {volume} {16}},\ \bibinfo {pages} {159} (\bibinfo {year} {2020})}\BibitemShut {NoStop}%
\bibitem [{\citenamefont {Zong}\ \emph {et~al.}(2021)\citenamefont {Zong}, \citenamefont {Dolgirev}, \citenamefont {Kogar}, \citenamefont {Su}, \citenamefont {Shen}, \citenamefont {Straquadine}, \citenamefont {Wang}, \citenamefont {Luo}, \citenamefont {Kozina}, \citenamefont {Reid}, \citenamefont {Li}, \citenamefont {Yang}, \citenamefont {Weathersby}, \citenamefont {Park}, \citenamefont {Sie}, \citenamefont {Jarillo-Herrero}, \citenamefont {Fisher}, \citenamefont {Wang}, \citenamefont {Demler},\ and\ \citenamefont {Gedik}}]{Zong2021RoleOrder}%
  \BibitemOpen
  \bibfield  {author} {\bibinfo {author} {\bibfnamefont {A.}~\bibnamefont {Zong}}, \bibinfo {author} {\bibfnamefont {P.~E.}\ \bibnamefont {Dolgirev}}, \bibinfo {author} {\bibfnamefont {A.}~\bibnamefont {Kogar}}, \bibinfo {author} {\bibfnamefont {Y.}~\bibnamefont {Su}}, \bibinfo {author} {\bibfnamefont {X.}~\bibnamefont {Shen}}, \bibinfo {author} {\bibfnamefont {J.~A.~W.}\ \bibnamefont {Straquadine}}, \bibinfo {author} {\bibfnamefont {X.}~\bibnamefont {Wang}}, \bibinfo {author} {\bibfnamefont {D.}~\bibnamefont {Luo}}, \bibinfo {author} {\bibfnamefont {M.~E.}\ \bibnamefont {Kozina}}, \bibinfo {author} {\bibfnamefont {A.~H.}\ \bibnamefont {Reid}}, \bibinfo {author} {\bibfnamefont {R.}~\bibnamefont {Li}}, \bibinfo {author} {\bibfnamefont {J.}~\bibnamefont {Yang}}, \bibinfo {author} {\bibfnamefont {S.~P.}\ \bibnamefont {Weathersby}}, \bibinfo {author} {\bibfnamefont {S.}~\bibnamefont {Park}}, \bibinfo {author} {\bibfnamefont {E.~J.}\ \bibnamefont {Sie}}, \bibinfo {author} {\bibfnamefont {P.}~\bibnamefont
  {Jarillo-Herrero}}, \bibinfo {author} {\bibfnamefont {I.~R.}\ \bibnamefont {Fisher}}, \bibinfo {author} {\bibfnamefont {X.}~\bibnamefont {Wang}}, \bibinfo {author} {\bibfnamefont {E.}~\bibnamefont {Demler}},\ and\ \bibinfo {author} {\bibfnamefont {N.}~\bibnamefont {Gedik}},\ }\bibfield  {title} {\bibinfo {title} {{Role of equilibrium fluctuations in light-induced order}},\ }\href {https://doi.org/10.1103/PhysRevLett.127.227401} {\bibfield  {journal} {\bibinfo  {journal} {Physical Review Letters}\ }\textbf {\bibinfo {volume} {127}},\ \bibinfo {pages} {227401} (\bibinfo {year} {2021})}\BibitemShut {NoStop}%
\bibitem [{\citenamefont {Cheng}\ \emph {et~al.}(2024)\citenamefont {Cheng}, \citenamefont {Zong}, \citenamefont {Wu}, \citenamefont {Meng}, \citenamefont {Xia}, \citenamefont {Qi}, \citenamefont {Zhu}, \citenamefont {Zou}, \citenamefont {Jiang}, \citenamefont {Guo}, \citenamefont {van Wezel}, \citenamefont {Kogar}, \citenamefont {Zuerch}, \citenamefont {Zhang}, \citenamefont {Zhu},\ and\ \citenamefont {Xiang}}]{Cheng2024UltrafastWave}%
  \BibitemOpen
  \bibfield  {author} {\bibinfo {author} {\bibfnamefont {Y.}~\bibnamefont {Cheng}}, \bibinfo {author} {\bibfnamefont {A.}~\bibnamefont {Zong}}, \bibinfo {author} {\bibfnamefont {L.}~\bibnamefont {Wu}}, \bibinfo {author} {\bibfnamefont {Q.}~\bibnamefont {Meng}}, \bibinfo {author} {\bibfnamefont {W.}~\bibnamefont {Xia}}, \bibinfo {author} {\bibfnamefont {F.}~\bibnamefont {Qi}}, \bibinfo {author} {\bibfnamefont {P.}~\bibnamefont {Zhu}}, \bibinfo {author} {\bibfnamefont {X.}~\bibnamefont {Zou}}, \bibinfo {author} {\bibfnamefont {T.}~\bibnamefont {Jiang}}, \bibinfo {author} {\bibfnamefont {Y.}~\bibnamefont {Guo}}, \bibinfo {author} {\bibfnamefont {J.}~\bibnamefont {van Wezel}}, \bibinfo {author} {\bibfnamefont {A.}~\bibnamefont {Kogar}}, \bibinfo {author} {\bibfnamefont {M.~W.}\ \bibnamefont {Zuerch}}, \bibinfo {author} {\bibfnamefont {J.}~\bibnamefont {Zhang}}, \bibinfo {author} {\bibfnamefont {Y.}~\bibnamefont {Zhu}},\ and\ \bibinfo {author} {\bibfnamefont {D.}~\bibnamefont {Xiang}},\ }\bibfield  {title}
  {\bibinfo {title} {{Ultrafast formation of topological defects in a two-dimensional charge density wave}},\ }\href {https://doi.org/10.1038/s41567-023-02279-x} {\bibfield  {journal} {\bibinfo  {journal} {Nature Physics}\ }\textbf {\bibinfo {volume} {20}},\ \bibinfo {pages} {54} (\bibinfo {year} {2024})}\BibitemShut {NoStop}%
\bibitem [{\citenamefont {Ravnik}\ \emph {et~al.}(2018)\citenamefont {Ravnik}, \citenamefont {Vaskivskyi}, \citenamefont {Mertelj},\ and\ \citenamefont {Mihailovic}}]{Ravnik2018}%
  \BibitemOpen
  \bibfield  {author} {\bibinfo {author} {\bibfnamefont {J.}~\bibnamefont {Ravnik}}, \bibinfo {author} {\bibfnamefont {I.}~\bibnamefont {Vaskivskyi}}, \bibinfo {author} {\bibfnamefont {T.}~\bibnamefont {Mertelj}},\ and\ \bibinfo {author} {\bibfnamefont {D.}~\bibnamefont {Mihailovic}},\ }\bibfield  {title} {\bibinfo {title} {{Real-time observation of the coherent transition to a metastable emergent state in 1T-TaS$_2$}},\ }\href {https://doi.org/10.1103/PhysRevB.97.075304} {\bibfield  {journal} {\bibinfo  {journal} {Physical Review B}\ }\textbf {\bibinfo {volume} {97}},\ \bibinfo {pages} {1} (\bibinfo {year} {2018})}\BibitemShut {NoStop}%
\bibitem [{\citenamefont {Vaskivskyi}\ \emph {et~al.}(2015)\citenamefont {Vaskivskyi}, \citenamefont {Gospodaric}, \citenamefont {Brazovskii}, \citenamefont {Svetin}, \citenamefont {Sutar}, \citenamefont {Goreshnik}, \citenamefont {Mihailovic}, \citenamefont {Mertelj},\ and\ \citenamefont {Mihailovic}}]{Vaskivskyi2015}%
  \BibitemOpen
  \bibfield  {author} {\bibinfo {author} {\bibfnamefont {I.}~\bibnamefont {Vaskivskyi}}, \bibinfo {author} {\bibfnamefont {J.}~\bibnamefont {Gospodaric}}, \bibinfo {author} {\bibfnamefont {S.}~\bibnamefont {Brazovskii}}, \bibinfo {author} {\bibfnamefont {D.}~\bibnamefont {Svetin}}, \bibinfo {author} {\bibfnamefont {P.}~\bibnamefont {Sutar}}, \bibinfo {author} {\bibfnamefont {E.}~\bibnamefont {Goreshnik}}, \bibinfo {author} {\bibfnamefont {I.~A.}\ \bibnamefont {Mihailovic}}, \bibinfo {author} {\bibfnamefont {T.}~\bibnamefont {Mertelj}},\ and\ \bibinfo {author} {\bibfnamefont {D.}~\bibnamefont {Mihailovic}},\ }\bibfield  {title} {\bibinfo {title} {{Controlling the metal-to-insulator relaxation of the metastable hidden quantum state in 1T-TaS$_2$}},\ }\href {https://doi.org/10.1126/sciadv.1500168} {\bibfield  {journal} {\bibinfo  {journal} {Science Advances}\ }\textbf {\bibinfo {volume} {1}},\ \bibinfo {pages} {e1500168} (\bibinfo {year} {2015})}\BibitemShut {NoStop}%
\bibitem [{\citenamefont {Mihailovic}(2019)}]{Mihailovic2019TheApplications}%
  \BibitemOpen
  \bibfield  {author} {\bibinfo {author} {\bibfnamefont {D.}~\bibnamefont {Mihailovic}},\ }\bibfield  {title} {\bibinfo {title} {{The importance of topological defects in photoexcited phase transitions including memory applications}},\ }\href {https://doi.org/10.3390/app9050890} {\bibfield  {journal} {\bibinfo  {journal} {Applied Sciences (Switzerland)}\ }\textbf {\bibinfo {volume} {9}},\ \bibinfo {pages} {890} (\bibinfo {year} {2019})}\BibitemShut {NoStop}%
\bibitem [{\citenamefont {Cheng}\ \emph {et~al.}(2022)\citenamefont {Cheng}, \citenamefont {Zong}, \citenamefont {Li}, \citenamefont {Xia}, \citenamefont {Duan}, \citenamefont {Zhao}, \citenamefont {Li}, \citenamefont {Qi}, \citenamefont {Wu}, \citenamefont {Zhao}, \citenamefont {Zhu}, \citenamefont {Zou}, \citenamefont {Jiang}, \citenamefont {Guo}, \citenamefont {Yang}, \citenamefont {Qian}, \citenamefont {Zhang}, \citenamefont {Kogar}, \citenamefont {Zuerch}, \citenamefont {Xiang},\ and\ \citenamefont {Zhang}}]{Cheng2022Light-inducedCorrelations}%
  \BibitemOpen
  \bibfield  {author} {\bibinfo {author} {\bibfnamefont {Y.}~\bibnamefont {Cheng}}, \bibinfo {author} {\bibfnamefont {A.}~\bibnamefont {Zong}}, \bibinfo {author} {\bibfnamefont {J.}~\bibnamefont {Li}}, \bibinfo {author} {\bibfnamefont {W.}~\bibnamefont {Xia}}, \bibinfo {author} {\bibfnamefont {S.}~\bibnamefont {Duan}}, \bibinfo {author} {\bibfnamefont {W.}~\bibnamefont {Zhao}}, \bibinfo {author} {\bibfnamefont {Y.}~\bibnamefont {Li}}, \bibinfo {author} {\bibfnamefont {F.}~\bibnamefont {Qi}}, \bibinfo {author} {\bibfnamefont {J.}~\bibnamefont {Wu}}, \bibinfo {author} {\bibfnamefont {L.}~\bibnamefont {Zhao}}, \bibinfo {author} {\bibfnamefont {P.}~\bibnamefont {Zhu}}, \bibinfo {author} {\bibfnamefont {X.}~\bibnamefont {Zou}}, \bibinfo {author} {\bibfnamefont {T.}~\bibnamefont {Jiang}}, \bibinfo {author} {\bibfnamefont {Y.}~\bibnamefont {Guo}}, \bibinfo {author} {\bibfnamefont {L.}~\bibnamefont {Yang}}, \bibinfo {author} {\bibfnamefont {D.}~\bibnamefont {Qian}}, \bibinfo {author} {\bibfnamefont {W.}~\bibnamefont
  {Zhang}}, \bibinfo {author} {\bibfnamefont {A.}~\bibnamefont {Kogar}}, \bibinfo {author} {\bibfnamefont {M.~W.}\ \bibnamefont {Zuerch}}, \bibinfo {author} {\bibfnamefont {D.}~\bibnamefont {Xiang}},\ and\ \bibinfo {author} {\bibfnamefont {J.}~\bibnamefont {Zhang}},\ }\bibfield  {title} {\bibinfo {title} {{Light-induced dimension crossover dictated by excitonic correlations}},\ }\href {https://doi.org/10.1038/s41467-022-28309-5} {\bibfield  {journal} {\bibinfo  {journal} {Nature Communications}\ }\textbf {\bibinfo {volume} {13}},\ \bibinfo {pages} {963} (\bibinfo {year} {2022})}\BibitemShut {NoStop}%
\bibitem [{\citenamefont {Ning}\ \emph {et~al.}(2024)\citenamefont {Ning}, \citenamefont {Oh}, \citenamefont {Su}, \citenamefont {von Hoegen}, \citenamefont {Porter}, \citenamefont {Capa~Salinas}, \citenamefont {Nguyen}, \citenamefont {Chollet}, \citenamefont {Sato}, \citenamefont {Esposito}, \citenamefont {Hoffmann}, \citenamefont {White}, \citenamefont {Melendrez}, \citenamefont {Zhu}, \citenamefont {Wilson},\ and\ \citenamefont {Gedik}}]{Ning2024DynamicalSuperconductor}%
  \BibitemOpen
  \bibfield  {author} {\bibinfo {author} {\bibfnamefont {H.}~\bibnamefont {Ning}}, \bibinfo {author} {\bibfnamefont {K.~H.}\ \bibnamefont {Oh}}, \bibinfo {author} {\bibfnamefont {Y.}~\bibnamefont {Su}}, \bibinfo {author} {\bibfnamefont {A.}~\bibnamefont {von Hoegen}}, \bibinfo {author} {\bibfnamefont {Z.}~\bibnamefont {Porter}}, \bibinfo {author} {\bibfnamefont {A.}~\bibnamefont {Capa~Salinas}}, \bibinfo {author} {\bibfnamefont {Q.~L.}\ \bibnamefont {Nguyen}}, \bibinfo {author} {\bibfnamefont {M.}~\bibnamefont {Chollet}}, \bibinfo {author} {\bibfnamefont {T.}~\bibnamefont {Sato}}, \bibinfo {author} {\bibfnamefont {V.}~\bibnamefont {Esposito}}, \bibinfo {author} {\bibfnamefont {M.~C.}\ \bibnamefont {Hoffmann}}, \bibinfo {author} {\bibfnamefont {A.}~\bibnamefont {White}}, \bibinfo {author} {\bibfnamefont {C.}~\bibnamefont {Melendrez}}, \bibinfo {author} {\bibfnamefont {D.}~\bibnamefont {Zhu}}, \bibinfo {author} {\bibfnamefont {S.~D.}\ \bibnamefont {Wilson}},\ and\ \bibinfo {author} {\bibfnamefont
  {N.}~\bibnamefont {Gedik}},\ }\bibfield  {title} {\bibinfo {title} {{Dynamical decoding of the competition between charge density waves in a kagome superconductor}},\ }\href {https://doi.org/10.1038/s41467-024-51485-5} {\bibfield  {journal} {\bibinfo  {journal} {Nature Communications}\ }\textbf {\bibinfo {volume} {15}},\ \bibinfo {pages} {7286} (\bibinfo {year} {2024})}\BibitemShut {NoStop}%
\bibitem [{\citenamefont {Su}\ \emph {et~al.}(2025)\citenamefont {Su}, \citenamefont {Lv}, \citenamefont {Zong}, \citenamefont {M{\"{u}}ller}, \citenamefont {Chattopadhyay}, \citenamefont {Dolgirev}, \citenamefont {Singh}, \citenamefont {Straquadine}, \citenamefont {Choi}, \citenamefont {Azoury}, \citenamefont {Mogi}, \citenamefont {Fisher}, \citenamefont {Demler},\ and\ \citenamefont {Gedik}}]{Su2025}%
  \BibitemOpen
  \bibfield  {author} {\bibinfo {author} {\bibfnamefont {Y.}~\bibnamefont {Su}}, \bibinfo {author} {\bibfnamefont {B.~Q.}\ \bibnamefont {Lv}}, \bibinfo {author} {\bibfnamefont {A.}~\bibnamefont {Zong}}, \bibinfo {author} {\bibfnamefont {A.}~\bibnamefont {M{\"{u}}ller}}, \bibinfo {author} {\bibfnamefont {S.}~\bibnamefont {Chattopadhyay}}, \bibinfo {author} {\bibfnamefont {P.~E.}\ \bibnamefont {Dolgirev}}, \bibinfo {author} {\bibfnamefont {A.~G.}\ \bibnamefont {Singh}}, \bibinfo {author} {\bibfnamefont {J.~A.~W.}\ \bibnamefont {Straquadine}}, \bibinfo {author} {\bibfnamefont {D.}~\bibnamefont {Choi}}, \bibinfo {author} {\bibfnamefont {D.}~\bibnamefont {Azoury}}, \bibinfo {author} {\bibfnamefont {M.}~\bibnamefont {Mogi}}, \bibinfo {author} {\bibfnamefont {I.~R.}\ \bibnamefont {Fisher}}, \bibinfo {author} {\bibfnamefont {E.}~\bibnamefont {Demler}},\ and\ \bibinfo {author} {\bibfnamefont {N.}~\bibnamefont {Gedik}},\ }\bibfield  {title} {\bibinfo {title} {{Time-domain identification of distinct mechanisms for
  competing charge density waves in a rare-earth tritelluride}},\ }\href {http://arxiv.org/abs/2503.13936} {\bibfield  {journal} {\bibinfo  {journal} {arXiv:2503.13936}\ } (\bibinfo {year} {2025})}\BibitemShut {NoStop}%
\bibitem [{\citenamefont {Overhauser}(1971)}]{Overhauser1971ObservabilityDiffraction}%
  \BibitemOpen
  \bibfield  {author} {\bibinfo {author} {\bibfnamefont {A.~W.}\ \bibnamefont {Overhauser}},\ }\bibfield  {title} {\bibinfo {title} {{Observability of charge-density waves by neutron diffraction}},\ }\href {https://doi.org/10.1103/PhysRevB.3.3173} {\bibfield  {journal} {\bibinfo  {journal} {Physical Review B}\ }\textbf {\bibinfo {volume} {3}},\ \bibinfo {pages} {3173} (\bibinfo {year} {1971})}\BibitemShut {NoStop}%
\bibitem [{\citenamefont {Rathore}\ \emph {et~al.}(2023)\citenamefont {Rathore}, \citenamefont {Pathak}, \citenamefont {Gupta}, \citenamefont {Mittal}, \citenamefont {Kulkarni}, \citenamefont {Thamizhavel}, \citenamefont {Singhal}, \citenamefont {Said},\ and\ \citenamefont {Bansal}}]{Rathore2023EvolutionEuTe4}%
  \BibitemOpen
  \bibfield  {author} {\bibinfo {author} {\bibfnamefont {R.}~\bibnamefont {Rathore}}, \bibinfo {author} {\bibfnamefont {A.}~\bibnamefont {Pathak}}, \bibinfo {author} {\bibfnamefont {M.~K.}\ \bibnamefont {Gupta}}, \bibinfo {author} {\bibfnamefont {R.}~\bibnamefont {Mittal}}, \bibinfo {author} {\bibfnamefont {R.}~\bibnamefont {Kulkarni}}, \bibinfo {author} {\bibfnamefont {A.}~\bibnamefont {Thamizhavel}}, \bibinfo {author} {\bibfnamefont {H.}~\bibnamefont {Singhal}}, \bibinfo {author} {\bibfnamefont {A.~H.}\ \bibnamefont {Said}},\ and\ \bibinfo {author} {\bibfnamefont {D.}~\bibnamefont {Bansal}},\ }\bibfield  {title} {\bibinfo {title} {{Evolution of static charge density wave order, amplitude mode dynamics, and suppression of Kohn anomalies at the hysteretic transition in EuTe$_4$}},\ }\href {https://doi.org/10.1103/PhysRevB.107.024101} {\bibfield  {journal} {\bibinfo  {journal} {Physical Review B}\ }\textbf {\bibinfo {volume} {107}},\ \bibinfo {pages} {024101} (\bibinfo {year} {2023})}\BibitemShut {NoStop}%
\bibitem [{\citenamefont {Ning}\ \emph {et~al.}(2025)\citenamefont {Ning}, \citenamefont {Oh}, \citenamefont {Su}, \citenamefont {Shi}, \citenamefont {Wu}, \citenamefont {Liu}, \citenamefont {Lv}, \citenamefont {Zong}, \citenamefont {Kang}, \citenamefont {Choi}, \citenamefont {Kim}, \citenamefont {Ha}, \citenamefont {Kim}, \citenamefont {Sarker}, \citenamefont {Ruff}, \citenamefont {Kim}, \citenamefont {Wang}, \citenamefont {Senthil}, \citenamefont {Jang},\ and\ \citenamefont {Gedik}}]{Ning2025}%
  \BibitemOpen
  \bibfield  {author} {\bibinfo {author} {\bibfnamefont {H.}~\bibnamefont {Ning}}, \bibinfo {author} {\bibfnamefont {K.~H.}\ \bibnamefont {Oh}}, \bibinfo {author} {\bibfnamefont {Y.}~\bibnamefont {Su}}, \bibinfo {author} {\bibfnamefont {Z.~D.}\ \bibnamefont {Shi}}, \bibinfo {author} {\bibfnamefont {D.}~\bibnamefont {Wu}}, \bibinfo {author} {\bibfnamefont {Q.}~\bibnamefont {Liu}}, \bibinfo {author} {\bibfnamefont {B.~Q.}\ \bibnamefont {Lv}}, \bibinfo {author} {\bibfnamefont {A.}~\bibnamefont {Zong}}, \bibinfo {author} {\bibfnamefont {G.}~\bibnamefont {Kang}}, \bibinfo {author} {\bibfnamefont {H.}~\bibnamefont {Choi}}, \bibinfo {author} {\bibfnamefont {H.-w.~J.}\ \bibnamefont {Kim}}, \bibinfo {author} {\bibfnamefont {S.}~\bibnamefont {Ha}}, \bibinfo {author} {\bibfnamefont {J.}~\bibnamefont {Kim}}, \bibinfo {author} {\bibfnamefont {S.}~\bibnamefont {Sarker}}, \bibinfo {author} {\bibfnamefont {J.~P.~C.}\ \bibnamefont {Ruff}}, \bibinfo {author} {\bibfnamefont {B.~J.}\ \bibnamefont {Kim}}, \bibinfo {author}
  {\bibfnamefont {N.~L.}\ \bibnamefont {Wang}}, \bibinfo {author} {\bibfnamefont {T.}~\bibnamefont {Senthil}}, \bibinfo {author} {\bibfnamefont {H.}~\bibnamefont {Jang}},\ and\ \bibinfo {author} {\bibfnamefont {N.}~\bibnamefont {Gedik}},\ }\bibfield  {title} {\bibinfo {title} {{Bidirectional Ultrafast Control of Charge Density Waves via Phase Competition}},\ }\href {https://doi.org/10.1103/b1vl-qlkk} {\bibfield  {journal} {\bibinfo  {journal} {Physical Review Letters}\ }\textbf {\bibinfo {volume} {135}},\ \bibinfo {pages} {246504} (\bibinfo {year} {2025})}\BibitemShut {NoStop}%
\bibitem [{\citenamefont {Lee}\ \emph {et~al.}(2012)\citenamefont {Lee}, \citenamefont {Chuang}, \citenamefont {Moore}, \citenamefont {Zhu}, \citenamefont {Patthey}, \citenamefont {Trigo}, \citenamefont {Lu}, \citenamefont {Kirchmann}, \citenamefont {Krupin}, \citenamefont {Yi}, \citenamefont {Langner}, \citenamefont {Huse}, \citenamefont {Robinson}, \citenamefont {Chen}, \citenamefont {Zhou}, \citenamefont {Coslovich}, \citenamefont {Huber}, \citenamefont {Reis}, \citenamefont {Kaindl}, \citenamefont {Schoenlein}, \citenamefont {Doering}, \citenamefont {Denes}, \citenamefont {Schlotter}, \citenamefont {Turner}, \citenamefont {Johnson}, \citenamefont {F{\"{o}}rst}, \citenamefont {Sasagawa}, \citenamefont {Kung}, \citenamefont {Sorini}, \citenamefont {Kemper}, \citenamefont {Moritz}, \citenamefont {Devereaux}, \citenamefont {Lee}, \citenamefont {Shen},\ and\ \citenamefont {Hussain}}]{Lee2012PhaseNickelate}%
  \BibitemOpen
  \bibfield  {author} {\bibinfo {author} {\bibfnamefont {W.~S.}\ \bibnamefont {Lee}}, \bibinfo {author} {\bibfnamefont {Y.~D.}\ \bibnamefont {Chuang}}, \bibinfo {author} {\bibfnamefont {R.~G.}\ \bibnamefont {Moore}}, \bibinfo {author} {\bibfnamefont {Y.}~\bibnamefont {Zhu}}, \bibinfo {author} {\bibfnamefont {L.}~\bibnamefont {Patthey}}, \bibinfo {author} {\bibfnamefont {M.}~\bibnamefont {Trigo}}, \bibinfo {author} {\bibfnamefont {D.~H.}\ \bibnamefont {Lu}}, \bibinfo {author} {\bibfnamefont {P.~S.}\ \bibnamefont {Kirchmann}}, \bibinfo {author} {\bibfnamefont {O.}~\bibnamefont {Krupin}}, \bibinfo {author} {\bibfnamefont {M.}~\bibnamefont {Yi}}, \bibinfo {author} {\bibfnamefont {M.}~\bibnamefont {Langner}}, \bibinfo {author} {\bibfnamefont {N.}~\bibnamefont {Huse}}, \bibinfo {author} {\bibfnamefont {J.~S.}\ \bibnamefont {Robinson}}, \bibinfo {author} {\bibfnamefont {Y.}~\bibnamefont {Chen}}, \bibinfo {author} {\bibfnamefont {S.~Y.}\ \bibnamefont {Zhou}}, \bibinfo {author} {\bibfnamefont {G.}~\bibnamefont
  {Coslovich}}, \bibinfo {author} {\bibfnamefont {B.}~\bibnamefont {Huber}}, \bibinfo {author} {\bibfnamefont {D.~A.}\ \bibnamefont {Reis}}, \bibinfo {author} {\bibfnamefont {R.~A.}\ \bibnamefont {Kaindl}}, \bibinfo {author} {\bibfnamefont {R.~W.}\ \bibnamefont {Schoenlein}}, \bibinfo {author} {\bibfnamefont {D.}~\bibnamefont {Doering}}, \bibinfo {author} {\bibfnamefont {P.}~\bibnamefont {Denes}}, \bibinfo {author} {\bibfnamefont {W.~F.}\ \bibnamefont {Schlotter}}, \bibinfo {author} {\bibfnamefont {J.~J.}\ \bibnamefont {Turner}}, \bibinfo {author} {\bibfnamefont {S.~L.}\ \bibnamefont {Johnson}}, \bibinfo {author} {\bibfnamefont {M.}~\bibnamefont {F{\"{o}}rst}}, \bibinfo {author} {\bibfnamefont {T.}~\bibnamefont {Sasagawa}}, \bibinfo {author} {\bibfnamefont {Y.~F.}\ \bibnamefont {Kung}}, \bibinfo {author} {\bibfnamefont {A.~P.}\ \bibnamefont {Sorini}}, \bibinfo {author} {\bibfnamefont {A.~F.}\ \bibnamefont {Kemper}}, \bibinfo {author} {\bibfnamefont {B.}~\bibnamefont {Moritz}}, \bibinfo {author} {\bibfnamefont
  {T.~P.}\ \bibnamefont {Devereaux}}, \bibinfo {author} {\bibfnamefont {D.~H.}\ \bibnamefont {Lee}}, \bibinfo {author} {\bibfnamefont {Z.~X.}\ \bibnamefont {Shen}},\ and\ \bibinfo {author} {\bibfnamefont {Z.}~\bibnamefont {Hussain}},\ }\bibfield  {title} {\bibinfo {title} {{Phase fluctuations and the absence of topological defects in a photo-excited charge-ordered nickelate}},\ }\href {https://doi.org/10.1038/ncomms1837} {\bibfield  {journal} {\bibinfo  {journal} {Nature Communications}\ }\textbf {\bibinfo {volume} {3}},\ \bibinfo {pages} {838} (\bibinfo {year} {2012})}\BibitemShut {NoStop}%
\bibitem [{\citenamefont {Chase}\ \emph {et~al.}(2016)\citenamefont {Chase}, \citenamefont {Trigo}, \citenamefont {Reid}, \citenamefont {Li}, \citenamefont {Vecchione}, \citenamefont {Shen}, \citenamefont {Weathersby}, \citenamefont {Coffee}, \citenamefont {Hartmann}, \citenamefont {Reis}, \citenamefont {Wang},\ and\ \citenamefont {D{\"{u}}rr}}]{Chase2016UltrafastFilms}%
  \BibitemOpen
  \bibfield  {author} {\bibinfo {author} {\bibfnamefont {T.}~\bibnamefont {Chase}}, \bibinfo {author} {\bibfnamefont {M.}~\bibnamefont {Trigo}}, \bibinfo {author} {\bibfnamefont {A.~H.}\ \bibnamefont {Reid}}, \bibinfo {author} {\bibfnamefont {R.}~\bibnamefont {Li}}, \bibinfo {author} {\bibfnamefont {T.}~\bibnamefont {Vecchione}}, \bibinfo {author} {\bibfnamefont {X.}~\bibnamefont {Shen}}, \bibinfo {author} {\bibfnamefont {S.}~\bibnamefont {Weathersby}}, \bibinfo {author} {\bibfnamefont {R.}~\bibnamefont {Coffee}}, \bibinfo {author} {\bibfnamefont {N.}~\bibnamefont {Hartmann}}, \bibinfo {author} {\bibfnamefont {D.~A.}\ \bibnamefont {Reis}}, \bibinfo {author} {\bibfnamefont {X.~J.}\ \bibnamefont {Wang}},\ and\ \bibinfo {author} {\bibfnamefont {H.~A.}\ \bibnamefont {D{\"{u}}rr}},\ }\bibfield  {title} {\bibinfo {title} {{Ultrafast electron diffraction from non-equilibrium phonons in femtosecond laser heated Au films}},\ }\href {https://doi.org/10.1063/1.4940981} {\bibfield  {journal} {\bibinfo  {journal} {Applied
  Physics Letters}\ }\textbf {\bibinfo {volume} {108}},\ \bibinfo {pages} {041909} (\bibinfo {year} {2016})}\BibitemShut {NoStop}%
\bibitem [{\citenamefont {Lin}\ \emph {et~al.}(2017)\citenamefont {Lin}, \citenamefont {Kochat}, \citenamefont {Krishnamoorthy}, \citenamefont {Bassman}, \citenamefont {Weninger}, \citenamefont {Zheng}, \citenamefont {Zhang}, \citenamefont {Apte}, \citenamefont {Tiwary}, \citenamefont {Shen}, \citenamefont {Li}, \citenamefont {Kalia}, \citenamefont {Ajayan}, \citenamefont {Nakano}, \citenamefont {Vashishta}, \citenamefont {Shimojo}, \citenamefont {Wang}, \citenamefont {Fritz},\ and\ \citenamefont {Bergmann}}]{Lin2017UltrafastMoSe2}%
  \BibitemOpen
  \bibfield  {author} {\bibinfo {author} {\bibfnamefont {M.~F.}\ \bibnamefont {Lin}}, \bibinfo {author} {\bibfnamefont {V.}~\bibnamefont {Kochat}}, \bibinfo {author} {\bibfnamefont {A.}~\bibnamefont {Krishnamoorthy}}, \bibinfo {author} {\bibfnamefont {L.}~\bibnamefont {Bassman}}, \bibinfo {author} {\bibfnamefont {C.}~\bibnamefont {Weninger}}, \bibinfo {author} {\bibfnamefont {Q.}~\bibnamefont {Zheng}}, \bibinfo {author} {\bibfnamefont {X.}~\bibnamefont {Zhang}}, \bibinfo {author} {\bibfnamefont {A.}~\bibnamefont {Apte}}, \bibinfo {author} {\bibfnamefont {C.~S.}\ \bibnamefont {Tiwary}}, \bibinfo {author} {\bibfnamefont {X.}~\bibnamefont {Shen}}, \bibinfo {author} {\bibfnamefont {R.}~\bibnamefont {Li}}, \bibinfo {author} {\bibfnamefont {R.}~\bibnamefont {Kalia}}, \bibinfo {author} {\bibfnamefont {P.}~\bibnamefont {Ajayan}}, \bibinfo {author} {\bibfnamefont {A.}~\bibnamefont {Nakano}}, \bibinfo {author} {\bibfnamefont {P.}~\bibnamefont {Vashishta}}, \bibinfo {author} {\bibfnamefont {F.}~\bibnamefont {Shimojo}},
  \bibinfo {author} {\bibfnamefont {X.}~\bibnamefont {Wang}}, \bibinfo {author} {\bibfnamefont {D.~M.}\ \bibnamefont {Fritz}},\ and\ \bibinfo {author} {\bibfnamefont {U.}~\bibnamefont {Bergmann}},\ }\bibfield  {title} {\bibinfo {title} {{Ultrafast non-radiative dynamics of atomically thin MoSe$_2$}},\ }\href {https://doi.org/10.1038/s41467-017-01844-2} {\bibfield  {journal} {\bibinfo  {journal} {Nature Communications}\ }\textbf {\bibinfo {volume} {8}},\ \bibinfo {pages} {1745} (\bibinfo {year} {2017})}\BibitemShut {NoStop}%
\bibitem [{\citenamefont {Tinnemann}\ \emph {et~al.}(2019)\citenamefont {Tinnemann}, \citenamefont {Streub{\"{u}}hr}, \citenamefont {Hafke}, \citenamefont {Kalus}, \citenamefont {Hanisch-Blicharski}, \citenamefont {Ligges}, \citenamefont {Zhou}, \citenamefont {Von Der~Linde}, \citenamefont {Bovensiepen},\ and\ \citenamefont {Horn-Von~Hoegen}}]{Tinnemann2019UltrafastTransfer}%
  \BibitemOpen
  \bibfield  {author} {\bibinfo {author} {\bibfnamefont {V.}~\bibnamefont {Tinnemann}}, \bibinfo {author} {\bibfnamefont {C.}~\bibnamefont {Streub{\"{u}}hr}}, \bibinfo {author} {\bibfnamefont {B.}~\bibnamefont {Hafke}}, \bibinfo {author} {\bibfnamefont {A.}~\bibnamefont {Kalus}}, \bibinfo {author} {\bibfnamefont {A.}~\bibnamefont {Hanisch-Blicharski}}, \bibinfo {author} {\bibfnamefont {M.}~\bibnamefont {Ligges}}, \bibinfo {author} {\bibfnamefont {P.}~\bibnamefont {Zhou}}, \bibinfo {author} {\bibfnamefont {D.}~\bibnamefont {Von Der~Linde}}, \bibinfo {author} {\bibfnamefont {U.}~\bibnamefont {Bovensiepen}},\ and\ \bibinfo {author} {\bibfnamefont {M.}~\bibnamefont {Horn-Von~Hoegen}},\ }\bibfield  {title} {\bibinfo {title} {{Ultrafast electron diffraction from a Bi(111) surface: Impulsive lattice excitation and Debye-Waller analysis at large momentum transfer}},\ }\href {https://doi.org/10.1063/1.5093637} {\bibfield  {journal} {\bibinfo  {journal} {Structural Dynamics}\ }\textbf {\bibinfo {volume} {6}},\ \bibinfo
  {pages} {035101} (\bibinfo {year} {2019})}\BibitemShut {NoStop}%
\bibitem [{\citenamefont {Gr{\"{u}}ner}(2018)}]{Gruner2018DensitySolids}%
  \BibitemOpen
  \bibfield  {author} {\bibinfo {author} {\bibfnamefont {G.}~\bibnamefont {Gr{\"{u}}ner}},\ }\href {https://doi.org/10.1201/9780429501012} {\emph {\bibinfo {title} {{Density Waves in Solids}}}}\ (\bibinfo  {publisher} {CRC Press},\ \bibinfo {year} {2018})\BibitemShut {NoStop}%
\bibitem [{\citenamefont {Sobota}\ \emph {et~al.}(2021)\citenamefont {Sobota}, \citenamefont {He},\ and\ \citenamefont {Shen}}]{Sobota2021Angle-resolvedMaterials}%
  \BibitemOpen
  \bibfield  {author} {\bibinfo {author} {\bibfnamefont {J.~A.}\ \bibnamefont {Sobota}}, \bibinfo {author} {\bibfnamefont {Y.}~\bibnamefont {He}},\ and\ \bibinfo {author} {\bibfnamefont {Z.-X.}\ \bibnamefont {Shen}},\ }\bibfield  {title} {\bibinfo {title} {{Angle-resolved photoemission studies of quantum materials}},\ }\href {https://doi.org/10.1103/RevModPhys.93.025006} {\bibfield  {journal} {\bibinfo  {journal} {Reviews of Modern Physics}\ }\textbf {\bibinfo {volume} {93}},\ \bibinfo {pages} {025006} (\bibinfo {year} {2021})}\BibitemShut {NoStop}%
\bibitem [{\citenamefont {Lv}\ \emph {et~al.}(2024)\citenamefont {Lv}, \citenamefont {Zong}, \citenamefont {Wu}, \citenamefont {Nie}, \citenamefont {Su}, \citenamefont {Choi}, \citenamefont {Ilyas}, \citenamefont {Fichera}, \citenamefont {Li}, \citenamefont {Baldini}, \citenamefont {Mogi}, \citenamefont {Huang}, \citenamefont {Po}, \citenamefont {Meng}, \citenamefont {Wang}, \citenamefont {Wang},\ and\ \citenamefont {Gedik}}]{Lv2024CoexistenceSemiconductor}%
  \BibitemOpen
  \bibfield  {author} {\bibinfo {author} {\bibfnamefont {B.}~\bibnamefont {Lv}}, \bibinfo {author} {\bibfnamefont {A.}~\bibnamefont {Zong}}, \bibinfo {author} {\bibfnamefont {D.}~\bibnamefont {Wu}}, \bibinfo {author} {\bibfnamefont {Z.}~\bibnamefont {Nie}}, \bibinfo {author} {\bibfnamefont {Y.}~\bibnamefont {Su}}, \bibinfo {author} {\bibfnamefont {D.}~\bibnamefont {Choi}}, \bibinfo {author} {\bibfnamefont {B.}~\bibnamefont {Ilyas}}, \bibinfo {author} {\bibfnamefont {B.~T.}\ \bibnamefont {Fichera}}, \bibinfo {author} {\bibfnamefont {J.}~\bibnamefont {Li}}, \bibinfo {author} {\bibfnamefont {E.}~\bibnamefont {Baldini}}, \bibinfo {author} {\bibfnamefont {M.}~\bibnamefont {Mogi}}, \bibinfo {author} {\bibfnamefont {Y.-B.}\ \bibnamefont {Huang}}, \bibinfo {author} {\bibfnamefont {H.~C.}\ \bibnamefont {Po}}, \bibinfo {author} {\bibfnamefont {S.}~\bibnamefont {Meng}}, \bibinfo {author} {\bibfnamefont {Y.}~\bibnamefont {Wang}}, \bibinfo {author} {\bibfnamefont {N.}~\bibnamefont {Wang}},\ and\ \bibinfo {author}
  {\bibfnamefont {N.}~\bibnamefont {Gedik}},\ }\bibfield  {title} {\bibinfo {title} {{Coexistence of interacting charge density waves in a layered semiconductor}},\ }\href {https://doi.org/10.1103/PhysRevLett.132.206401} {\bibfield  {journal} {\bibinfo  {journal} {Physical Review Letters}\ }\textbf {\bibinfo {volume} {132}},\ \bibinfo {pages} {206401} (\bibinfo {year} {2024})}\BibitemShut {NoStop}%
\bibitem [{\citenamefont {Rischel}\ \emph {et~al.}(1997)\citenamefont {Rischel}, \citenamefont {Rousse}, \citenamefont {Uschmann}, \citenamefont {Albouy}, \citenamefont {Geindre}, \citenamefont {Audebert}, \citenamefont {Gauthier}, \citenamefont {Fr{\"{o}}ster}, \citenamefont {Martin},\ and\ \citenamefont {Antonetti}}]{Rischel1997FemtosecondFilms}%
  \BibitemOpen
  \bibfield  {author} {\bibinfo {author} {\bibfnamefont {C.}~\bibnamefont {Rischel}}, \bibinfo {author} {\bibfnamefont {A.}~\bibnamefont {Rousse}}, \bibinfo {author} {\bibfnamefont {I.}~\bibnamefont {Uschmann}}, \bibinfo {author} {\bibfnamefont {P.-A.}\ \bibnamefont {Albouy}}, \bibinfo {author} {\bibfnamefont {J.-P.}\ \bibnamefont {Geindre}}, \bibinfo {author} {\bibfnamefont {P.}~\bibnamefont {Audebert}}, \bibinfo {author} {\bibfnamefont {J.-C.}\ \bibnamefont {Gauthier}}, \bibinfo {author} {\bibfnamefont {E.}~\bibnamefont {Fr{\"{o}}ster}}, \bibinfo {author} {\bibfnamefont {J.-L.}\ \bibnamefont {Martin}},\ and\ \bibinfo {author} {\bibfnamefont {A.}~\bibnamefont {Antonetti}},\ }\bibfield  {title} {\bibinfo {title} {{Femtosecond time-resolved X-ray diffraction from laser-heated organic films}},\ }\href {https://doi.org/10.1038/37317} {\bibfield  {journal} {\bibinfo  {journal} {Nature}\ }\textbf {\bibinfo {volume} {390}},\ \bibinfo {pages} {490} (\bibinfo {year} {1997})}\BibitemShut {NoStop}%
\bibitem [{\citenamefont {Beaud}\ \emph {et~al.}(2014)\citenamefont {Beaud}, \citenamefont {Caviezel}, \citenamefont {Mariager}, \citenamefont {Rettig}, \citenamefont {Ingold}, \citenamefont {Dornes}, \citenamefont {Huang}, \citenamefont {Johnson}, \citenamefont {Radovic}, \citenamefont {Huber}, \citenamefont {Kubacka}, \citenamefont {Ferrer}, \citenamefont {Lemke}, \citenamefont {Chollet}, \citenamefont {Zhu}, \citenamefont {Glownia}, \citenamefont {Sikorski}, \citenamefont {Robert}, \citenamefont {Wadati}, \citenamefont {Nakamura}, \citenamefont {Kawasaki}, \citenamefont {Tokura}, \citenamefont {Johnson},\ and\ \citenamefont {Staub}}]{Beaud2014ATransitions}%
  \BibitemOpen
  \bibfield  {author} {\bibinfo {author} {\bibfnamefont {P.}~\bibnamefont {Beaud}}, \bibinfo {author} {\bibfnamefont {A.}~\bibnamefont {Caviezel}}, \bibinfo {author} {\bibfnamefont {S.~O.}\ \bibnamefont {Mariager}}, \bibinfo {author} {\bibfnamefont {L.}~\bibnamefont {Rettig}}, \bibinfo {author} {\bibfnamefont {G.}~\bibnamefont {Ingold}}, \bibinfo {author} {\bibfnamefont {C.}~\bibnamefont {Dornes}}, \bibinfo {author} {\bibfnamefont {S.~W.}\ \bibnamefont {Huang}}, \bibinfo {author} {\bibfnamefont {J.~A.}\ \bibnamefont {Johnson}}, \bibinfo {author} {\bibfnamefont {M.}~\bibnamefont {Radovic}}, \bibinfo {author} {\bibfnamefont {T.}~\bibnamefont {Huber}}, \bibinfo {author} {\bibfnamefont {T.}~\bibnamefont {Kubacka}}, \bibinfo {author} {\bibfnamefont {A.}~\bibnamefont {Ferrer}}, \bibinfo {author} {\bibfnamefont {H.~T.}\ \bibnamefont {Lemke}}, \bibinfo {author} {\bibfnamefont {M.}~\bibnamefont {Chollet}}, \bibinfo {author} {\bibfnamefont {D.}~\bibnamefont {Zhu}}, \bibinfo {author} {\bibfnamefont {J.~M.}\ \bibnamefont
  {Glownia}}, \bibinfo {author} {\bibfnamefont {M.}~\bibnamefont {Sikorski}}, \bibinfo {author} {\bibfnamefont {A.}~\bibnamefont {Robert}}, \bibinfo {author} {\bibfnamefont {H.}~\bibnamefont {Wadati}}, \bibinfo {author} {\bibfnamefont {M.}~\bibnamefont {Nakamura}}, \bibinfo {author} {\bibfnamefont {M.}~\bibnamefont {Kawasaki}}, \bibinfo {author} {\bibfnamefont {Y.}~\bibnamefont {Tokura}}, \bibinfo {author} {\bibfnamefont {S.~L.}\ \bibnamefont {Johnson}},\ and\ \bibinfo {author} {\bibfnamefont {U.}~\bibnamefont {Staub}},\ }\bibfield  {title} {\bibinfo {title} {{A time-dependent order parameter for ultrafast photoinduced phase transitions}},\ }\href {https://doi.org/10.1038/nmat4046} {\bibfield  {journal} {\bibinfo  {journal} {Nature Materials}\ }\textbf {\bibinfo {volume} {13}},\ \bibinfo {pages} {923} (\bibinfo {year} {2014})}\BibitemShut {NoStop}%
\bibitem [{\citenamefont {Banerjee}\ \emph {et~al.}(2013)\citenamefont {Banerjee}, \citenamefont {Feng}, \citenamefont {Silevitch}, \citenamefont {Wang}, \citenamefont {Lang}, \citenamefont {Kuo}, \citenamefont {Fisher},\ and\ \citenamefont {Rosenbaum}}]{Banerjee2013ChargeTellurides}%
  \BibitemOpen
  \bibfield  {author} {\bibinfo {author} {\bibfnamefont {A.}~\bibnamefont {Banerjee}}, \bibinfo {author} {\bibfnamefont {Y.}~\bibnamefont {Feng}}, \bibinfo {author} {\bibfnamefont {D.~M.}\ \bibnamefont {Silevitch}}, \bibinfo {author} {\bibfnamefont {J.}~\bibnamefont {Wang}}, \bibinfo {author} {\bibfnamefont {J.~C.}\ \bibnamefont {Lang}}, \bibinfo {author} {\bibfnamefont {H.~H.}\ \bibnamefont {Kuo}}, \bibinfo {author} {\bibfnamefont {I.~R.}\ \bibnamefont {Fisher}},\ and\ \bibinfo {author} {\bibfnamefont {T.~F.}\ \bibnamefont {Rosenbaum}},\ }\bibfield  {title} {\bibinfo {title} {{Charge transfer and multiple density waves in the rare earth tellurides}},\ }\href {https://doi.org/10.1103/PhysRevB.87.155131} {\bibfield  {journal} {\bibinfo  {journal} {Physical Review B}\ }\textbf {\bibinfo {volume} {87}},\ \bibinfo {pages} {155131} (\bibinfo {year} {2013})}\BibitemShut {NoStop}%
\bibitem [{\citenamefont {Moore}\ \emph {et~al.}(2016)\citenamefont {Moore}, \citenamefont {Lee}, \citenamefont {Kirchman}, \citenamefont {Chuang}, \citenamefont {Kemper}, \citenamefont {Trigo}, \citenamefont {Patthey}, \citenamefont {Lu}, \citenamefont {Krupin}, \citenamefont {Yi}, \citenamefont {Reis}, \citenamefont {Doering}, \citenamefont {Denes}, \citenamefont {Schlotter}, \citenamefont {Turner}, \citenamefont {Hays}, \citenamefont {Hering}, \citenamefont {Benson}, \citenamefont {Chu}, \citenamefont {Devereaux}, \citenamefont {Fisher}, \citenamefont {Hussain},\ and\ \citenamefont {Shen}}]{Moore2016Ultrafast3}%
  \BibitemOpen
  \bibfield  {author} {\bibinfo {author} {\bibfnamefont {R.~G.}\ \bibnamefont {Moore}}, \bibinfo {author} {\bibfnamefont {W.~S.}\ \bibnamefont {Lee}}, \bibinfo {author} {\bibfnamefont {P.~S.}\ \bibnamefont {Kirchman}}, \bibinfo {author} {\bibfnamefont {Y.~D.}\ \bibnamefont {Chuang}}, \bibinfo {author} {\bibfnamefont {A.~F.}\ \bibnamefont {Kemper}}, \bibinfo {author} {\bibfnamefont {M.}~\bibnamefont {Trigo}}, \bibinfo {author} {\bibfnamefont {L.}~\bibnamefont {Patthey}}, \bibinfo {author} {\bibfnamefont {D.~H.}\ \bibnamefont {Lu}}, \bibinfo {author} {\bibfnamefont {O.}~\bibnamefont {Krupin}}, \bibinfo {author} {\bibfnamefont {M.}~\bibnamefont {Yi}}, \bibinfo {author} {\bibfnamefont {D.~A.}\ \bibnamefont {Reis}}, \bibinfo {author} {\bibfnamefont {D.}~\bibnamefont {Doering}}, \bibinfo {author} {\bibfnamefont {P.}~\bibnamefont {Denes}}, \bibinfo {author} {\bibfnamefont {W.~F.}\ \bibnamefont {Schlotter}}, \bibinfo {author} {\bibfnamefont {J.~J.}\ \bibnamefont {Turner}}, \bibinfo {author} {\bibfnamefont
  {G.}~\bibnamefont {Hays}}, \bibinfo {author} {\bibfnamefont {P.}~\bibnamefont {Hering}}, \bibinfo {author} {\bibfnamefont {T.}~\bibnamefont {Benson}}, \bibinfo {author} {\bibfnamefont {J.-H.}\ \bibnamefont {Chu}}, \bibinfo {author} {\bibfnamefont {T.~P.}\ \bibnamefont {Devereaux}}, \bibinfo {author} {\bibfnamefont {I.~R.}\ \bibnamefont {Fisher}}, \bibinfo {author} {\bibfnamefont {Z.}~\bibnamefont {Hussain}},\ and\ \bibinfo {author} {\bibfnamefont {Z.-X.}\ \bibnamefont {Shen}},\ }\bibfield  {title} {\bibinfo {title} {{Ultrafast resonant soft x-ray diffraction dynamics of the charge density wave in TbTe$_3$}},\ }\href {https://doi.org/10.1103/PhysRevB.93.024304} {\bibfield  {journal} {\bibinfo  {journal} {Physical Review B}\ }\textbf {\bibinfo {volume} {93}},\ \bibinfo {pages} {024304} (\bibinfo {year} {2016})}\BibitemShut {NoStop}%
\bibitem [{\citenamefont {Blanco-Canosa}\ \emph {et~al.}(2014)\citenamefont {Blanco-Canosa}, \citenamefont {Frano}, \citenamefont {Schierle}, \citenamefont {Porras}, \citenamefont {Loew}, \citenamefont {Minola}, \citenamefont {Bluschke}, \citenamefont {Weschke}, \citenamefont {Keimer},\ and\ \citenamefont {Le~Tacon}}]{Blanco-Canosa2014ResonantX}%
  \BibitemOpen
  \bibfield  {author} {\bibinfo {author} {\bibfnamefont {S.}~\bibnamefont {Blanco-Canosa}}, \bibinfo {author} {\bibfnamefont {A.}~\bibnamefont {Frano}}, \bibinfo {author} {\bibfnamefont {E.}~\bibnamefont {Schierle}}, \bibinfo {author} {\bibfnamefont {J.}~\bibnamefont {Porras}}, \bibinfo {author} {\bibfnamefont {T.}~\bibnamefont {Loew}}, \bibinfo {author} {\bibfnamefont {M.}~\bibnamefont {Minola}}, \bibinfo {author} {\bibfnamefont {M.}~\bibnamefont {Bluschke}}, \bibinfo {author} {\bibfnamefont {E.}~\bibnamefont {Weschke}}, \bibinfo {author} {\bibfnamefont {B.}~\bibnamefont {Keimer}},\ and\ \bibinfo {author} {\bibfnamefont {M.}~\bibnamefont {Le~Tacon}},\ }\bibfield  {title} {\bibinfo {title} {{Resonant x-ray scattering study of charge-density wave correlations in YBa$_{2}$Cu$_{3}$O$_{6+x}$}},\ }\href {https://doi.org/10.1103/PhysRevB.90.054513} {\bibfield  {journal} {\bibinfo  {journal} {Physical Review B}\ }\textbf {\bibinfo {volume} {90}},\ \bibinfo {pages} {054513} (\bibinfo {year} {2014})}\BibitemShut
  {NoStop}%
\bibitem [{\citenamefont {Jang}\ \emph {et~al.}(2016)\citenamefont {Jang}, \citenamefont {Lee}, \citenamefont {Nojiri}, \citenamefont {Matsuzawa}, \citenamefont {Yasumura}, \citenamefont {Nie}, \citenamefont {Maharaj}, \citenamefont {Gerber}, \citenamefont {Liu}, \citenamefont {Mehta}, \citenamefont {Bonn}, \citenamefont {Liang}, \citenamefont {Hardy}, \citenamefont {Burns}, \citenamefont {Islam}, \citenamefont {Song}, \citenamefont {Hastings}, \citenamefont {Devereaux}, \citenamefont {Shen}, \citenamefont {Kivelson}, \citenamefont {Kao}, \citenamefont {Zhu},\ and\ \citenamefont {Lee}}]{Jang2016IdealYBCO}%
  \BibitemOpen
  \bibfield  {author} {\bibinfo {author} {\bibfnamefont {H.}~\bibnamefont {Jang}}, \bibinfo {author} {\bibfnamefont {W.~S.}\ \bibnamefont {Lee}}, \bibinfo {author} {\bibfnamefont {H.}~\bibnamefont {Nojiri}}, \bibinfo {author} {\bibfnamefont {S.}~\bibnamefont {Matsuzawa}}, \bibinfo {author} {\bibfnamefont {H.}~\bibnamefont {Yasumura}}, \bibinfo {author} {\bibfnamefont {L.}~\bibnamefont {Nie}}, \bibinfo {author} {\bibfnamefont {A.~V.}\ \bibnamefont {Maharaj}}, \bibinfo {author} {\bibfnamefont {S.}~\bibnamefont {Gerber}}, \bibinfo {author} {\bibfnamefont {Y.~J.}\ \bibnamefont {Liu}}, \bibinfo {author} {\bibfnamefont {A.}~\bibnamefont {Mehta}}, \bibinfo {author} {\bibfnamefont {D.~A.}\ \bibnamefont {Bonn}}, \bibinfo {author} {\bibfnamefont {R.}~\bibnamefont {Liang}}, \bibinfo {author} {\bibfnamefont {W.~N.}\ \bibnamefont {Hardy}}, \bibinfo {author} {\bibfnamefont {C.~A.}\ \bibnamefont {Burns}}, \bibinfo {author} {\bibfnamefont {Z.}~\bibnamefont {Islam}}, \bibinfo {author} {\bibfnamefont {S.}~\bibnamefont {Song}},
  \bibinfo {author} {\bibfnamefont {J.}~\bibnamefont {Hastings}}, \bibinfo {author} {\bibfnamefont {T.~P.}\ \bibnamefont {Devereaux}}, \bibinfo {author} {\bibfnamefont {Z.~X.}\ \bibnamefont {Shen}}, \bibinfo {author} {\bibfnamefont {S.~A.}\ \bibnamefont {Kivelson}}, \bibinfo {author} {\bibfnamefont {C.~C.}\ \bibnamefont {Kao}}, \bibinfo {author} {\bibfnamefont {D.}~\bibnamefont {Zhu}},\ and\ \bibinfo {author} {\bibfnamefont {J.~S.}\ \bibnamefont {Lee}},\ }\bibfield  {title} {\bibinfo {title} {{Ideal charge-density-wave order in the high-field state of superconducting YBCO}},\ }\href {https://doi.org/10.1073/pnas.1612849113} {\bibfield  {journal} {\bibinfo  {journal} {Proc. Natl. Acad. Sci. U.S.A.}\ }\textbf {\bibinfo {volume} {113}},\ \bibinfo {pages} {14645} (\bibinfo {year} {2016})}\BibitemShut {NoStop}%
\bibitem [{\citenamefont {Jacques}\ \emph {et~al.}(2016)\citenamefont {Jacques}, \citenamefont {Laulh{\'{e}}}, \citenamefont {Moisan}, \citenamefont {Ravy},\ and\ \citenamefont {Le~Bolloc’h}}]{Jacques2016Laser-InducedChromium}%
  \BibitemOpen
  \bibfield  {author} {\bibinfo {author} {\bibfnamefont {V.}~\bibnamefont {Jacques}}, \bibinfo {author} {\bibfnamefont {C.}~\bibnamefont {Laulh{\'{e}}}}, \bibinfo {author} {\bibfnamefont {N.}~\bibnamefont {Moisan}}, \bibinfo {author} {\bibfnamefont {S.}~\bibnamefont {Ravy}},\ and\ \bibinfo {author} {\bibfnamefont {D.}~\bibnamefont {Le~Bolloc’h}},\ }\bibfield  {title} {\bibinfo {title} {{Laser-Induced Charge-Density-Wave Transient Depinning in Chromium}},\ }\href {https://doi.org/10.1103/PhysRevLett.117.156401} {\bibfield  {journal} {\bibinfo  {journal} {Physical Review Letters}\ }\textbf {\bibinfo {volume} {117}},\ \bibinfo {pages} {156401} (\bibinfo {year} {2016})}\BibitemShut {NoStop}%
\bibitem [{\citenamefont {Zhou}\ \emph {et~al.}(2021)\citenamefont {Zhou}, \citenamefont {Williams}, \citenamefont {Sun}, \citenamefont {Malliakas}, \citenamefont {Kanatzidis}, \citenamefont {Kemper},\ and\ \citenamefont {Ruan}}]{Zhou2021NonequilibriumWave}%
  \BibitemOpen
  \bibfield  {author} {\bibinfo {author} {\bibfnamefont {F.}~\bibnamefont {Zhou}}, \bibinfo {author} {\bibfnamefont {J.}~\bibnamefont {Williams}}, \bibinfo {author} {\bibfnamefont {S.}~\bibnamefont {Sun}}, \bibinfo {author} {\bibfnamefont {C.~D.}\ \bibnamefont {Malliakas}}, \bibinfo {author} {\bibfnamefont {M.~G.}\ \bibnamefont {Kanatzidis}}, \bibinfo {author} {\bibfnamefont {A.~F.}\ \bibnamefont {Kemper}},\ and\ \bibinfo {author} {\bibfnamefont {C.~Y.}\ \bibnamefont {Ruan}},\ }\bibfield  {title} {\bibinfo {title} {{Nonequilibrium dynamics of spontaneous symmetry breaking into a hidden state of charge-density wave}},\ }\bibfield  {journal} {\bibinfo  {journal} {Nature Communications}\ }\textbf {\bibinfo {volume} {12}},\ \href {https://doi.org/10.1038/s41467-020-20834-5} {10.1038/s41467-020-20834-5} (\bibinfo {year} {2021})\BibitemShut {NoStop}%
\bibitem [{\citenamefont {Vogelgesang}\ \emph {et~al.}(2018)\citenamefont {Vogelgesang}, \citenamefont {Storeck}, \citenamefont {Horstmann}, \citenamefont {Diekmann}, \citenamefont {Sivis}, \citenamefont {Schramm}, \citenamefont {Rossnagel}, \citenamefont {Sch{\"{a}}fer},\ and\ \citenamefont {Ropers}}]{Vogelgesang2018PhaseDiffraction}%
  \BibitemOpen
  \bibfield  {author} {\bibinfo {author} {\bibfnamefont {S.}~\bibnamefont {Vogelgesang}}, \bibinfo {author} {\bibfnamefont {G.}~\bibnamefont {Storeck}}, \bibinfo {author} {\bibfnamefont {J.~G.}\ \bibnamefont {Horstmann}}, \bibinfo {author} {\bibfnamefont {T.}~\bibnamefont {Diekmann}}, \bibinfo {author} {\bibfnamefont {M.}~\bibnamefont {Sivis}}, \bibinfo {author} {\bibfnamefont {S.}~\bibnamefont {Schramm}}, \bibinfo {author} {\bibfnamefont {K.}~\bibnamefont {Rossnagel}}, \bibinfo {author} {\bibfnamefont {S.}~\bibnamefont {Sch{\"{a}}fer}},\ and\ \bibinfo {author} {\bibfnamefont {C.}~\bibnamefont {Ropers}},\ }\bibfield  {title} {\bibinfo {title} {{Phase ordering of charge density waves traced by ultrafast low-energy electron diffraction}},\ }\href {https://doi.org/10.1038/nphys4309} {\bibfield  {journal} {\bibinfo  {journal} {Nature Physics}\ }\textbf {\bibinfo {volume} {14}},\ \bibinfo {pages} {184} (\bibinfo {year} {2018})}\BibitemShut {NoStop}%
\bibitem [{\citenamefont {Wang}\ and\ \citenamefont {Wang}(2019)}]{Wang2019WritingPulses}%
  \BibitemOpen
  \bibfield  {author} {\bibinfo {author} {\bibfnamefont {Y.}~\bibnamefont {Wang}}\ and\ \bibinfo {author} {\bibfnamefont {D.}~\bibnamefont {Wang}},\ }\bibfield  {title} {\bibinfo {title} {{Writing and erasing topological defects in charge density wave materials with femtosecond laser pulses}},\ }\href {https://doi.org/10.1364/ol.44.002939} {\bibfield  {journal} {\bibinfo  {journal} {Optics Letters}\ }\textbf {\bibinfo {volume} {44}},\ \bibinfo {pages} {2939} (\bibinfo {year} {2019})}\BibitemShut {NoStop}%
\bibitem [{\citenamefont {Weathersby}\ \emph {et~al.}(2015)\citenamefont {Weathersby}, \citenamefont {Brown}, \citenamefont {Centurion}, \citenamefont {Chase}, \citenamefont {Coffee}, \citenamefont {Corbett}, \citenamefont {Eichner}, \citenamefont {Frisch}, \citenamefont {Fry}, \citenamefont {G{\"{u}}hr}, \citenamefont {Hartmann}, \citenamefont {Hast}, \citenamefont {Hettel}, \citenamefont {Jobe}, \citenamefont {Jongewaard}, \citenamefont {Lewandowski}, \citenamefont {Li}, \citenamefont {Lindenberg}, \citenamefont {Makasyuk}, \citenamefont {May}, \citenamefont {McCormick}, \citenamefont {Nguyen}, \citenamefont {Reid}, \citenamefont {Shen}, \citenamefont {Sokolowski-Tinten}, \citenamefont {Vecchione}, \citenamefont {Vetter}, \citenamefont {Wu}, \citenamefont {Yang}, \citenamefont {D{\"{u}}rr},\ and\ \citenamefont {Wang}}]{Weathersby2015Mega-electron-voltLaboratory}%
  \BibitemOpen
  \bibfield  {author} {\bibinfo {author} {\bibfnamefont {S.~P.}\ \bibnamefont {Weathersby}}, \bibinfo {author} {\bibfnamefont {G.}~\bibnamefont {Brown}}, \bibinfo {author} {\bibfnamefont {M.}~\bibnamefont {Centurion}}, \bibinfo {author} {\bibfnamefont {T.~F.}\ \bibnamefont {Chase}}, \bibinfo {author} {\bibfnamefont {R.}~\bibnamefont {Coffee}}, \bibinfo {author} {\bibfnamefont {J.}~\bibnamefont {Corbett}}, \bibinfo {author} {\bibfnamefont {J.~P.}\ \bibnamefont {Eichner}}, \bibinfo {author} {\bibfnamefont {J.~C.}\ \bibnamefont {Frisch}}, \bibinfo {author} {\bibfnamefont {A.~R.}\ \bibnamefont {Fry}}, \bibinfo {author} {\bibfnamefont {M.}~\bibnamefont {G{\"{u}}hr}}, \bibinfo {author} {\bibfnamefont {N.}~\bibnamefont {Hartmann}}, \bibinfo {author} {\bibfnamefont {C.}~\bibnamefont {Hast}}, \bibinfo {author} {\bibfnamefont {R.}~\bibnamefont {Hettel}}, \bibinfo {author} {\bibfnamefont {R.~K.}\ \bibnamefont {Jobe}}, \bibinfo {author} {\bibfnamefont {E.~N.}\ \bibnamefont {Jongewaard}}, \bibinfo {author} {\bibfnamefont
  {J.~R.}\ \bibnamefont {Lewandowski}}, \bibinfo {author} {\bibfnamefont {R.~K.}\ \bibnamefont {Li}}, \bibinfo {author} {\bibfnamefont {A.~M.}\ \bibnamefont {Lindenberg}}, \bibinfo {author} {\bibfnamefont {I.}~\bibnamefont {Makasyuk}}, \bibinfo {author} {\bibfnamefont {J.~E.}\ \bibnamefont {May}}, \bibinfo {author} {\bibfnamefont {D.}~\bibnamefont {McCormick}}, \bibinfo {author} {\bibfnamefont {M.~N.}\ \bibnamefont {Nguyen}}, \bibinfo {author} {\bibfnamefont {A.~H.}\ \bibnamefont {Reid}}, \bibinfo {author} {\bibfnamefont {X.}~\bibnamefont {Shen}}, \bibinfo {author} {\bibfnamefont {K.}~\bibnamefont {Sokolowski-Tinten}}, \bibinfo {author} {\bibfnamefont {T.}~\bibnamefont {Vecchione}}, \bibinfo {author} {\bibfnamefont {S.~L.}\ \bibnamefont {Vetter}}, \bibinfo {author} {\bibfnamefont {J.}~\bibnamefont {Wu}}, \bibinfo {author} {\bibfnamefont {J.}~\bibnamefont {Yang}}, \bibinfo {author} {\bibfnamefont {H.~A.}\ \bibnamefont {D{\"{u}}rr}},\ and\ \bibinfo {author} {\bibfnamefont {X.~J.}\ \bibnamefont {Wang}},\
  }\bibfield  {title} {\bibinfo {title} {{Mega-electron-volt ultrafast electron diffraction at SLAC National Accelerator Laboratory}},\ }\href {https://doi.org/10.1063/1.4926994} {\bibfield  {journal} {\bibinfo  {journal} {Review of Scientific Instruments}\ }\textbf {\bibinfo {volume} {86}},\ \bibinfo {pages} {073702} (\bibinfo {year} {2015})}\BibitemShut {NoStop}%
\bibitem [{\citenamefont {Jang}\ \emph {et~al.}(2020)\citenamefont {Jang}, \citenamefont {Kim}, \citenamefont {Kim}, \citenamefont {Park}, \citenamefont {Kwon}, \citenamefont {Lee}, \citenamefont {Park}, \citenamefont {Park}, \citenamefont {Kim}, \citenamefont {Hyun}, \citenamefont {Hwang}, \citenamefont {Lee}, \citenamefont {Lim}, \citenamefont {Gang}, \citenamefont {Kim}, \citenamefont {Heo}, \citenamefont {Kim}, \citenamefont {Jung}, \citenamefont {Kim}, \citenamefont {Park}, \citenamefont {Kim}, \citenamefont {Shin}, \citenamefont {Park}, \citenamefont {Koo}, \citenamefont {Shin}, \citenamefont {Heo}, \citenamefont {Kim}, \citenamefont {Min}, \citenamefont {Han}, \citenamefont {Kang}, \citenamefont {Lee}, \citenamefont {Kim}, \citenamefont {Eom},\ and\ \citenamefont {Rah}}]{Jang2020Time-resolvedLaser}%
  \BibitemOpen
  \bibfield  {author} {\bibinfo {author} {\bibfnamefont {H.}~\bibnamefont {Jang}}, \bibinfo {author} {\bibfnamefont {H.~D.}\ \bibnamefont {Kim}}, \bibinfo {author} {\bibfnamefont {M.}~\bibnamefont {Kim}}, \bibinfo {author} {\bibfnamefont {S.~H.}\ \bibnamefont {Park}}, \bibinfo {author} {\bibfnamefont {S.}~\bibnamefont {Kwon}}, \bibinfo {author} {\bibfnamefont {J.~Y.}\ \bibnamefont {Lee}}, \bibinfo {author} {\bibfnamefont {S.~Y.}\ \bibnamefont {Park}}, \bibinfo {author} {\bibfnamefont {G.}~\bibnamefont {Park}}, \bibinfo {author} {\bibfnamefont {S.}~\bibnamefont {Kim}}, \bibinfo {author} {\bibfnamefont {H.}~\bibnamefont {Hyun}}, \bibinfo {author} {\bibfnamefont {S.}~\bibnamefont {Hwang}}, \bibinfo {author} {\bibfnamefont {C.~S.}\ \bibnamefont {Lee}}, \bibinfo {author} {\bibfnamefont {C.~Y.}\ \bibnamefont {Lim}}, \bibinfo {author} {\bibfnamefont {W.}~\bibnamefont {Gang}}, \bibinfo {author} {\bibfnamefont {M.}~\bibnamefont {Kim}}, \bibinfo {author} {\bibfnamefont {S.}~\bibnamefont {Heo}}, \bibinfo {author}
  {\bibfnamefont {J.}~\bibnamefont {Kim}}, \bibinfo {author} {\bibfnamefont {G.}~\bibnamefont {Jung}}, \bibinfo {author} {\bibfnamefont {S.}~\bibnamefont {Kim}}, \bibinfo {author} {\bibfnamefont {J.}~\bibnamefont {Park}}, \bibinfo {author} {\bibfnamefont {J.}~\bibnamefont {Kim}}, \bibinfo {author} {\bibfnamefont {H.}~\bibnamefont {Shin}}, \bibinfo {author} {\bibfnamefont {J.}~\bibnamefont {Park}}, \bibinfo {author} {\bibfnamefont {T.~Y.}\ \bibnamefont {Koo}}, \bibinfo {author} {\bibfnamefont {H.~J.}\ \bibnamefont {Shin}}, \bibinfo {author} {\bibfnamefont {H.}~\bibnamefont {Heo}}, \bibinfo {author} {\bibfnamefont {C.}~\bibnamefont {Kim}}, \bibinfo {author} {\bibfnamefont {C.~K.}\ \bibnamefont {Min}}, \bibinfo {author} {\bibfnamefont {J.~H.}\ \bibnamefont {Han}}, \bibinfo {author} {\bibfnamefont {H.~S.}\ \bibnamefont {Kang}}, \bibinfo {author} {\bibfnamefont {H.~S.}\ \bibnamefont {Lee}}, \bibinfo {author} {\bibfnamefont {K.~S.}\ \bibnamefont {Kim}}, \bibinfo {author} {\bibfnamefont {I.}~\bibnamefont {Eom}},\
  and\ \bibinfo {author} {\bibfnamefont {S.}~\bibnamefont {Rah}},\ }\bibfield  {title} {\bibinfo {title} {{Time-resolved resonant elastic soft x-ray scattering at Pohang Accelerator Laboratory X-ray Free Electron Laser}},\ }\href {https://doi.org/10.1063/5.0016414} {\bibfield  {journal} {\bibinfo  {journal} {Review of Scientific Instruments}\ }\textbf {\bibinfo {volume} {91}},\ \bibinfo {pages} {083904} (\bibinfo {year} {2020})}\BibitemShut {NoStop}%
\bibitem [{\citenamefont {Freelon}\ \emph {et~al.}(2023)\citenamefont {Freelon}, \citenamefont {Rohwer}, \citenamefont {Zong}, \citenamefont {Kogar}, \citenamefont {Zhou}, \citenamefont {Wong}, \citenamefont {Erge{\c{c}}en},\ and\ \citenamefont {Gedik}}]{Freelon2023DesignInstrument}%
  \BibitemOpen
  \bibfield  {author} {\bibinfo {author} {\bibfnamefont {B.}~\bibnamefont {Freelon}}, \bibinfo {author} {\bibfnamefont {T.}~\bibnamefont {Rohwer}}, \bibinfo {author} {\bibfnamefont {A.}~\bibnamefont {Zong}}, \bibinfo {author} {\bibfnamefont {A.}~\bibnamefont {Kogar}}, \bibinfo {author} {\bibfnamefont {H.}~\bibnamefont {Zhou}}, \bibinfo {author} {\bibfnamefont {L.~J.}\ \bibnamefont {Wong}}, \bibinfo {author} {\bibfnamefont {E.}~\bibnamefont {Erge{\c{c}}en}},\ and\ \bibinfo {author} {\bibfnamefont {N.}~\bibnamefont {Gedik}},\ }\bibfield  {title} {\bibinfo {title} {{Design and construction of a compact, high-repetition-rate ultrafast electron diffraction instrument}},\ }\href {https://doi.org/10.1063/5.0094278} {\bibfield  {journal} {\bibinfo  {journal} {Review of Scientific Instruments}\ }\textbf {\bibinfo {volume} {94}},\ \bibinfo {pages} {053305} (\bibinfo {year} {2023})}\BibitemShut {NoStop}%
\bibitem [{\citenamefont {Su}\ \emph {et~al.}(2023)\citenamefont {Su}, \citenamefont {Zong}, \citenamefont {Kogar}, \citenamefont {Lu}, \citenamefont {Hong}, \citenamefont {Freelon}, \citenamefont {Rohwer}, \citenamefont {Wang}, \citenamefont {Hwang},\ and\ \citenamefont {Gedik}}]{Su2023}%
  \BibitemOpen
  \bibfield  {author} {\bibinfo {author} {\bibfnamefont {Y.}~\bibnamefont {Su}}, \bibinfo {author} {\bibfnamefont {A.}~\bibnamefont {Zong}}, \bibinfo {author} {\bibfnamefont {A.}~\bibnamefont {Kogar}}, \bibinfo {author} {\bibfnamefont {D.}~\bibnamefont {Lu}}, \bibinfo {author} {\bibfnamefont {S.~S.}\ \bibnamefont {Hong}}, \bibinfo {author} {\bibfnamefont {B.}~\bibnamefont {Freelon}}, \bibinfo {author} {\bibfnamefont {T.}~\bibnamefont {Rohwer}}, \bibinfo {author} {\bibfnamefont {B.~Y.}\ \bibnamefont {Wang}}, \bibinfo {author} {\bibfnamefont {H.~Y.}\ \bibnamefont {Hwang}},\ and\ \bibinfo {author} {\bibfnamefont {N.}~\bibnamefont {Gedik}},\ }\bibfield  {title} {\bibinfo {title} {{Delamination-Assisted Ultrafast Wrinkle Formation in a Freestanding Film}},\ }\href {https://doi.org/10.1021/acs.nanolett.3c02898} {\bibfield  {journal} {\bibinfo  {journal} {Nano Letters}\ }\textbf {\bibinfo {volume} {23}},\ \bibinfo {pages} {10772} (\bibinfo {year} {2023})},\ \Eprint {https://arxiv.org/abs/2212.12082} {2212.12082}
  \BibitemShut {NoStop}%
\bibitem [{\citenamefont {Lee}\ \emph {et~al.}(2020)\citenamefont {Lee}, \citenamefont {Rohwer}, \citenamefont {Sie}, \citenamefont {Zong}, \citenamefont {Baldini}, \citenamefont {Straquadine}, \citenamefont {Walmsley}, \citenamefont {Gardner}, \citenamefont {Lee}, \citenamefont {Fisher},\ and\ \citenamefont {Gedik}}]{Lee2020HighPulses}%
  \BibitemOpen
  \bibfield  {author} {\bibinfo {author} {\bibfnamefont {C.}~\bibnamefont {Lee}}, \bibinfo {author} {\bibfnamefont {T.}~\bibnamefont {Rohwer}}, \bibinfo {author} {\bibfnamefont {E.~J.}\ \bibnamefont {Sie}}, \bibinfo {author} {\bibfnamefont {A.}~\bibnamefont {Zong}}, \bibinfo {author} {\bibfnamefont {E.}~\bibnamefont {Baldini}}, \bibinfo {author} {\bibfnamefont {J.}~\bibnamefont {Straquadine}}, \bibinfo {author} {\bibfnamefont {P.}~\bibnamefont {Walmsley}}, \bibinfo {author} {\bibfnamefont {D.}~\bibnamefont {Gardner}}, \bibinfo {author} {\bibfnamefont {Y.~S.}\ \bibnamefont {Lee}}, \bibinfo {author} {\bibfnamefont {I.~R.}\ \bibnamefont {Fisher}},\ and\ \bibinfo {author} {\bibfnamefont {N.}~\bibnamefont {Gedik}},\ }\bibfield  {title} {\bibinfo {title} {{High resolution time- And angle-resolved photoemission spectroscopy with 11 eV laser pulses}},\ }\href {https://doi.org/10.1063/1.5139556} {\bibfield  {journal} {\bibinfo  {journal} {Review of Scientific Instruments}\ }\textbf {\bibinfo {volume} {91}},\ \bibinfo
  {pages} {043102} (\bibinfo {year} {2020})}\BibitemShut {NoStop}%
\bibitem [{\citenamefont {Sie}\ \emph {et~al.}(2019)\citenamefont {Sie}, \citenamefont {Rohwer}, \citenamefont {Lee},\ and\ \citenamefont {Gedik}}]{Sie2019Time-resolvedXUVResolution}%
  \BibitemOpen
  \bibfield  {author} {\bibinfo {author} {\bibfnamefont {E.~J.}\ \bibnamefont {Sie}}, \bibinfo {author} {\bibfnamefont {T.}~\bibnamefont {Rohwer}}, \bibinfo {author} {\bibfnamefont {C.}~\bibnamefont {Lee}},\ and\ \bibinfo {author} {\bibfnamefont {N.}~\bibnamefont {Gedik}},\ }\bibfield  {title} {\bibinfo {title} {{Time-resolved XUV ARPES with tunable 24-33 eV laser pulses at 30 meV resolution}},\ }\href {https://doi.org/10.1038/s41467-019-11492-3} {\bibfield  {journal} {\bibinfo  {journal} {Nature Communications}\ }\textbf {\bibinfo {volume} {10}},\ \bibinfo {pages} {3535} (\bibinfo {year} {2019})}\BibitemShut {NoStop}%
\end{thebibliography}
\end{document}